# Continuous atomic displacements and lattice distortion during fcc-bcc martensitic transformation.

Cyril Cayron

Laboratory of ThermoMechanical Metallurgy (LMTM), PX Group Chair, Ecole Polytechnique Fédérale de Lausanne (EPFL), Rue de la Maladière 71b, 2000 Neuchâtel, Switzerland.
cyril.cayron@epfl.ch

**Abstract**   The continuous matrices of atomic displacements and lattice distortion from face-centred-cubic (fcc) to body-centred-cubic (bcc) phases compatible with the hard-sphere geometry of iron atoms are calculated for different possible final orientation relationships, such as Bain, Pitsch and Kurdjumov-Sachs (KS). The angular distortion introduced in the calculations appears as a natural order parameter of the transition. The distortion matrix associated to KS is assumed to represent the natural transformation mechanism; it has special mathematical properties and seems sufficient to explain the {225} habit planes with a simple criterion. From these calculations, the fcc-bcc martensitic transformation appears to be of type "angular distortive", and not "shear" as usually assumed.

**1. Introduction**

Martensitic transformations between face centred cubic (γ, fcc) austenite and body centred cubic (α, bcc) martensite have been studied for more than one century. Among the earlier works, one can cite those of Adolf Martens [1] and Floris Osmond [2] linked to the development of metallography. In 1924, Bain [3] proposed a simple model of fcc-bcc transformation. Few years later, the orientation relationship (OR) between austenite and martensite was measured by X-ray diffraction by Kurdjumov & Sachs [4], Nishiyama [5], and Wassermann [6]. The KS and NW ORs are separated by 5°, and both are at 10° far from the Bain OR. The discrepancy between the "expected" Bain OR and the experimental ORs made these authors propose separately a similar model of lattice distortion by shear and dilatation, which is known as the KSN model. This approach does not seem to have convinced the scientists of that time because they continued their efforts to conciliate the theoretical Bain distortion with the experimentally determined ORs and the shapes of the martensite laths, following Greninger and Troiano's approach [7]. This led in the 1950s to the phenomenological theory of martensite transformation/crystallography (PTMT or PTMC) [8].

The PTMC treats the lattice deformations, shape and internal defects of the martensite product in a coherent entangled way. Two important considerations are in the core of this theory: a) martensitic transformations occur according to shear mechanisms, and b) the Bain distortion has the lowest strains. Details of the theory can be found in many reference books such as those of Nishiyama [9], Christian [10] and Bhadeshia [11]. The PTMC is now applied to a large family of phase transition materials beyond the historical steels and iron alloys, such as TiNi and other shape-memory alloys [12]. In steels, the PTMC nicely describes the $\{259\}_\gamma$ and $\{3\ 10\ 15\}_\gamma$ habit planes, but more complex calculations are required for the $\{225\}_\gamma$ habit planes [9][13][14]. Despite its success and its board use, PTMC remains phenomenological. The exact atomic displacements of the iron atoms during the transformation are not in its scope: "*The crystallographic theory of martensite is on the hand called phenomenological; the steps into which the transformation is factorised are not unique and do not necessarily describe the actual path by which the atoms move from one lattice to the other. The theory simply provides a definite link between the initial and final states without being certain of the path in between*" [11].

Actually, neither the Bain distortion with PTMC nor the KSN model take into account the atomic movements; they are pure lattice transformation models. To our knowledge, Bogers and Burgers in 1964 [15] were the first to consider the iron atoms as hard spheres, and they proposes lattice deformations that are compatible with the atom sizes. Their approach, later refined by Olson and Cohen [16] assumes that the transformation results from a (multi)shear mechanism, but is not based on the Bain distortion. For the few last years, we have also tried to develop a model that could describe the atomic displacements during the transformation. Our motivation dates from our observation of some peculiar features in the pole figures obtained by X-ray diffraction and Electron BackScatter Diffraction (EBSD); they are spread between KS and the other classical ORs such as NW, Pitsch and Greninger-Troiano (GT) and could be simulated by two continuous rotations **A** and **B** [17]. Since these features are observed in many steels, iron-nickel alloys and meteorites, and brass (independently of the morphology and habit planes), we considered them as the trace of the plastic accommodation of the lattice deformation. In 2013, we proposed a hard-sphere model of fcc-bcc martensitic transformations [18], which is based on the Pitsch OR [19]. We called it "one-step" in comparison with a "two-step" model published few years earlier in which we imagined that the transformation was produced by the movement of partial Shockley dislocations leading to an intermediate fleeting hcp phase [17]. In the one-step Pitsch model the continuous features in the pole figures could be explained by taking into account the deformation field imposed by the transformation in the fcc surrounding matrix. Some experimental observations, such as the nucleation of the martensite at intersection of stacking faults planes were qualitatively described.

The present work is in the continuity of the one-step model [17]. The calculations rely on pure geometry and have no ambition to compete with the PTMC. As schematically represented in Fig. 1, PTMC treats the lattice deformations, shape and internal defects of the martensite product in a coherent way, whereas our approach only tries to determine the lattice deformations compatible with a hard-sphere rule, i.e. avoiding the interpenetration of the iron atoms during the lattice distortion. Our approach relies on an assumption different from PTMC. Instead of considering that the natural distortion occurs by Bain and that the final OR results from an accommodation process necessary to establish an invariant plane strain; we consider that the natural distortion is actually the one that gives the final OR and that the accommodation in the austenitic matrix is only a consequence of that distortion. Many questions could be raised from this approach: a) what is the mathematical expression of the distortions that lead to Pitsch and KS ORs, b) what is their link with the Bain distortion, c) is there a possibility to explain the habit planes from this approach? It will be shown that the distortion matrix associated to KS OR could constitute the natural distortion of the fcc-bcc martensitic transformation; it has very special properties and seems sufficient to explain the $\{225\}_\gamma$ habit planes.

## 2. Notations, composition rules of matrices and conventions

Before entering into the details of the calculations, it is worth clarifying the notations and conventions that will be used in the paper. The vectors are in bold small letters, such as **u** for vectors of direct space and **g** for vectors of reciprocal space. The matrices are in bold capital letters, such as the distortion matrices **D**. The bases are indicated by the letter **B**. A matrix of change of coordinates between two bases $\mathbf{B}_i$ and $\mathbf{B}_j$ is noted [$\mathbf{B}_i \rightarrow \mathbf{B}_j$]. Its three columns are constituted of the basic vectors of $\mathbf{B}_j$ written in the basis $\mathbf{B}_i$. A vector **u** written in a basis $\mathbf{B}_j$ is noted $\mathbf{u}_j$. The same vector expressed in the basis $\mathbf{B}_i$ is given by $\mathbf{u}_i = [\mathbf{B}_i \rightarrow \mathbf{B}_j] \mathbf{u}_j$. In a crystal of phase γ, noted by the same letter for sake of simplicity, the reference orthonormal basis formed by the vectors $[100]_\gamma$, $[010]_\gamma$, $[001]_\gamma$ is written $\mathbf{B}_0^\gamma = (\mathbf{a}^\gamma, \mathbf{b}^\gamma, \mathbf{c}^\gamma)$. Any vector of the crystal γ can be expressed in another basis $\mathbf{B}_i^\gamma = (\mathbf{x}^\gamma, \mathbf{y}^\gamma, \mathbf{z}^\gamma)$ with the help of the coordinate-transformation matrix $[\mathbf{B}_0^\gamma \rightarrow \mathbf{B}_i^\gamma] = \mathbf{B}_i^\gamma$. In the 2D case of Fig. 2a, it is

$$\mathbf{B}_i^\gamma = \begin{bmatrix} 1/2 & -1/2 \\ 1/2 & 1/2 \end{bmatrix} \quad (1)$$

When two crystals of phases γ and α are in orientation relationship, the parallelism conditions can be expressed by the existence of a common basis $\mathbf{B}_c$ constituted by three normalized vectors (not necessarily orthonormal). The coordinate transformation matrix is then given by $[\mathbf{B}_0^\gamma \to \mathbf{B}_c][\mathbf{B}_c \to \mathbf{B}_0^\alpha]$. For example, in the 2D case of the Fig. 2b, the common basis between the square and rectangular crystals is $\mathbf{B}_c = \left(\frac{\mathbf{x}^\gamma}{\|\mathbf{x}^\gamma\|}, \frac{\mathbf{y}^\gamma}{\|\mathbf{y}^\gamma\|}\right) = \left(\frac{\mathbf{a}^\alpha}{\|\mathbf{a}^\alpha\|}, \frac{\mathbf{b}^\alpha}{\|\mathbf{b}^\alpha\|}\right)$. If the unity is attributed the lattice parameter $a_\gamma$, and if $\mathbf{a}^\alpha = \mathbf{x}^\gamma$ and $\mathbf{b}^\alpha = k\,\mathbf{y}^\gamma$, the coordinate transformation matrix from crystal $\gamma$ to crystal $\alpha$ is

$$[\mathbf{B}_0^\gamma \to \mathbf{B}_0^\alpha] = \begin{bmatrix} 1/2 & -1/2 \\ 1/2 & 1/2 \end{bmatrix}\begin{bmatrix} 1 & 0 \\ 0 & k \end{bmatrix} = \begin{bmatrix} 1/2 & -k/2 \\ 1/2 & k/2 \end{bmatrix} \quad (2)$$

A homogeneous lattice distortion is a linear transformation **D** from the lattice $\gamma$ to the lattice $\gamma'$. Since **D** transforms the lattice $\mathbf{B}_0^\gamma = (\mathbf{a}^\gamma, \mathbf{b}^\gamma, \mathbf{c}^\gamma)$ into the distorted basis $\mathbf{B}_0^{\gamma'} = (\mathbf{a}^{\gamma'}, \mathbf{b}^{\gamma'}, \mathbf{c}^{\gamma'})$, it can be expressed in the initial basis by $\mathbf{D}_0 = [\mathbf{B}_0^\gamma \to \mathbf{B}_0^{\gamma'}]$. It can also be written in another basis of the initial crystal $\mathbf{B}_i^\gamma$ by $\mathbf{D}_i = [\mathbf{B}_i^\gamma \to \mathbf{B}_0^\gamma]\,\mathbf{D}_0\,[\mathbf{B}_0^\gamma \to \mathbf{B}_i^\gamma] = (\mathbf{B}_i^\gamma)^{-1}\,\mathbf{D}_0\,\mathbf{B}_i^\gamma$. One should be very careful to avoid any confusion with coordinate transformation matrices. A distortion matrices is "active", it changes any vector **u** of the initial crystal $\gamma$ into a new vector **u'** = **D.u**, whereas a coordinate transformation matrix is "passive", it allows to calculate the coordinate of a fixed vector **u** in different bases and tells nothing on the distortion mechanism.

We would like to introduce here a class of transformations that, as far as we know, is not reported in the textbooks but that will be very important for the rest of the paper. We call them "angular distortive" transformations. In such transformations, a vector is let invariant and another one is rotated by an angle $\theta$ without modification of its length, as represented in Fig. 2c. When the distortion applies to the vectors of the basis $\mathbf{B}_0$, the distortion matrix is

$$\mathbf{D}_0(\theta) = \begin{bmatrix} 1 & \sin(\theta) \\ 0 & \cos(\theta) \end{bmatrix} \quad (3)$$

This type of distortion is different from the usual simple shear transformation of amplitude s represented in Fig. 2d, even if a dilatation parameter $\delta$ perpendicular to the invariant plane is introduced. The expression of an invariant plane strain matrix is

$$\mathbf{D}_0(\theta) = \begin{bmatrix} 1 & s \\ 0 & 1+\delta \end{bmatrix} \quad (4)$$

and can be assimilated to the "angular distortive" matrix (3) only for special combination of s and $\delta \leq 1$. It will be shown in the rest of paper that angular distortive transformations are the ideal tool to treat lattice distortions that respect the hard sphere packing of the atoms.

The displacement field **F** associated to a homogeneous distortion **D** is **F.x** = (**I-D**).**x** with **I** the identity matrix. The gradient of displacements is simply $d\mathbf{F}/d\mathbf{x} = \mathbf{I-D}$.

For convenience in the calculations, the unity will be attributed to the lattice parameter of the fcc phase. The angles are given in radians, except those indicated in degrees (°). The numerical calculations and 3D graphics and surfaces are obtained with Mathematica 10. As in ref. [18], it is assumed in first and rough approximation that during the γ →α martensitic transformation, the iron atoms are hard spheres of same diameter in both phases, which implies that:

$$\sqrt{2}\, a_\gamma = \sqrt{3}\, a_\alpha \tag{5}$$

**3. Model of atomic displacements with Bain OR**

The Bain distortion is a contraction of 20% along one $<100>_\gamma$ axis and dilatations along the two $<011>_\gamma$ axis perpendicular to the contraction axis. The final Bain OR is $[100]_\gamma$ // $[100]_\alpha$, $[011]_\gamma$ // $[010]_\alpha$ and $[0\bar{1}1]_\gamma$ // $[001]_\alpha$. To our knowledge, the exact link between the continuous contraction and dilatation values has never been established. That is very surprising because the calculations are straightforward when assuming that the iron atoms are hard spheres that "roll" on each other during the distortion. In order to obtain a coherency with the reference frames used with our previous paper [18], we will take for convention that the contraction occurs along the **a**-axis (and not along the **c**-axis as usually chosen). Let us consider the basis $\mathbf{B}_1$ = (**x**, **y**, **z**) such that the **x**, **y** and **z** axes are orientated along the $[110]_\gamma$ and $[\bar{1}10]_\gamma$ and $[001]_\gamma$ axes, respectively. During the Bain distortion the **x** and **y** directions are rotated in opposite directions, as shown in Fig. 3a. Let us call α/2 the semi-angle of rotation of **x** and **y**. By projection, it is easy to show that the deformation (contraction) along **a** is given by $\Delta a/a = \sqrt{2}\cos(\pi/4 + \alpha/2) = \sqrt{2}\sin(\pi/4 - \alpha/2)$ and the resulting deformation (dilatation) along **b** is given by $\Delta b/b = \sqrt{2}\sin(\pi/4 + \alpha/2)$. Since the atoms along the $[100]_\gamma$ direction are shared by the (**a**, **b**) = $(001)_\gamma$ and (**a**, **c**) = $(010)_\gamma$ planes, the resulting deformation (dilatation) along **c** is the same as the one calculated along **b**, as shown in Fig. 4. Therefore, the Bain distortion matrix in the reference basis $\mathbf{B}_0$ is

$$\mathbf{D}_0^{Bain}(\alpha) = \sqrt{2}\begin{bmatrix} \sin(\pi/4 - \alpha/2) & 0 & 0 \\ 0 & \sin(\pi/4 + \alpha/2) & 0 \\ 0 & 0 & \sin(\pi/4 + \alpha/2) \end{bmatrix} \tag{6}$$

The transformation starts with the angle α = 0 and $\mathbf{D}_0^{Bain}(0)$ is the identity matrix. During the distortion, the $[110]_\gamma$ and $[\bar{1}10]_\gamma$ axes which make an angle of π/2 are transformed into the $[\bar{1}11]_\alpha$ and $[1\bar{1}1]_\alpha$ axes which make an angle of arccos(1/3). The angular difference is π/2-arccos(1/3) = arcsin(1/3). Thus, the fcc-bcc transformation is completed when the rotation angle reaches $\alpha_{max}$ = arcsin(1/3). By using trigonometric relations and simplification of radical forms, it can be shown that

$$\begin{cases} \sin(\frac{\alpha_{max}}{2}) = \frac{\sqrt{2}-1}{\sqrt{6}} \\ \cos(\frac{\alpha_{max}}{2}) = \frac{1}{\sqrt{6}.(\sqrt{2}-1)} \end{cases} \quad (7)$$

Therefore, the distortion matrix of the complete Bain transformation is

$$\mathbf{D}_0^{Bain} = \begin{bmatrix} \cos(\frac{\alpha_{max}}{2}) - \sin(\frac{\alpha_{max}}{2}) & 0 & 0 \\ 0 & \cos(\frac{\alpha_{max}}{2}) + \sin(\frac{\alpha_{max}}{2}) & 0 \\ 0 & 0 & \cos(\frac{\alpha_{max}}{2}) + \sin(\frac{\alpha_{max}}{2}) \end{bmatrix} \quad (8)$$

$$= \begin{bmatrix} \sqrt{\frac{2}{3}} & 0 & 0 \\ 0 & \frac{2}{\sqrt{3}} & 0 \\ 0 & 0 & \frac{2}{\sqrt{3}} \end{bmatrix} = \begin{bmatrix} 0.81 & 0 & 0 \\ 0 & 1.15 & 0 \\ 0 & 0 & 1.15 \end{bmatrix}$$

This gives the classical values of Bain deformation if one assumes that the atom size remains unchanged during the transformation, i.e. a compression of 18.3% along the **a**-axis and a dilatation of 15.5% along the **b** and **c** axes.

### 4. Model of atomic displacements with Pitsch OR

The Pitsch OR is $[110]_\gamma // [111]_\alpha$, $[\bar{1}10]_\gamma // [11\bar{2}]_\alpha$ and $[001]_\gamma // [\bar{1}10]_\alpha$. These axes form an orthogonal (but not orthonormal) reference basis $\mathbf{B}_1 = (\mathbf{x}, \mathbf{y}, \mathbf{z})$. During the Pitsch distortion, the axis **x** remains invariant, i.e. undistorted and unrotated; we called it "neutral line" [18] (the **x** direction is inverted in comparison to ref. [18] in order to work in right-hand basis). The **y** axis is rotated by an angle α; and the **z** axis remains parallel to the **c** axis and is elongated in order to respect the hard-sphere packing of the iron atoms, as shown in Fig. 3b. There are two ways to calculate the distortion. The first one uses Fig. 5 and the calculations are performed in a Cartesian coordinate system. The distance PM = QM = 1 becomes after deformation PM' = $\sqrt{1-\sin(\alpha)}$ and QM' = $\sqrt{1+\sin(\alpha)}$. Thus:

$$\Delta z/z = \Delta y/y = \frac{QM'}{QM} = \sqrt{1+\sin(\alpha)} \quad (9)$$

The second way consists in noticing that the elongation along **z** with Pitsch distortion should be equal to the one obtained with Bain because both models are based on hard-sphere packing, and thus:

$$\Delta z/z = \sqrt{2} \sin(\pi/4 + \alpha/2) \quad (10)$$

Trigonometric relations show that both equations (9) and (10) are equal.

Hence, the Pitsch distortion matrix written in the reference basis $\mathbf{B}_1$ is:

$$\mathbf{D}_1^{Pitsch}(\alpha) = \begin{bmatrix} 1 & \sin(\alpha) & 0 \\ 0 & \cos(\alpha) & 0 \\ 0 & 0 & \sqrt{1+\sin(\alpha)} \end{bmatrix} \quad (11)$$

In the reference basis $\mathbf{B}_0$, the basis $\mathbf{B}_1$ is given by the coordinate transformation matrix:

$$[\mathbf{B}_0 \to \mathbf{B}_1] = \begin{bmatrix} 1/2 & -1/2 & 0 \\ 1/2 & 1/2 & 0 \\ 0 & 0 & 1 \end{bmatrix}, \text{ for which the inverse is } [\mathbf{B}_1 \to \mathbf{B}_0] = \begin{bmatrix} 1 & 1 & 0 \\ -1 & 1 & 0 \\ 0 & 0 & 1 \end{bmatrix}$$

The Pitsch distortion matrix in the basis $B_0$ is therefore:

$$\mathbf{D}_0^{Pitsch}(\alpha) = [\mathbf{B}_0 \to \mathbf{B}_1]\mathbf{D}_1[\mathbf{B}_1 \to \mathbf{B}_0] \quad (12)$$

which becomes after calculations:

$$\mathbf{D}_0^{Pitsch}(\alpha) = \begin{bmatrix} \dfrac{1-\sin(\alpha)+\cos(\alpha)}{2} & \dfrac{1+\sin(\alpha)-\cos(\alpha)}{2} & 0 \\ \dfrac{1-\sin(\alpha)-\cos(\alpha)}{2} & \dfrac{1+\sin(\alpha)+\cos(\alpha)}{2} & 0 \\ 0 & 0 & \sqrt{1+\sin(\alpha)} \end{bmatrix} \quad (13)$$

It can be equivalently written:

$$\mathbf{D}_0^{Pitsch}(\alpha) = \begin{bmatrix} \dfrac{1+\sqrt{2}\cos(\pi/4+\alpha)}{2} & \dfrac{1-\sqrt{2}\cos(\pi/4+\alpha)}{2} & 0 \\ \dfrac{1-\sqrt{2}\cos(\pi/4-\alpha)}{2} & \dfrac{1+\sqrt{2}\cos(\pi/4-\alpha)}{2} & 0 \\ 0 & 0 & \sqrt{2}\sin(\pi/4+\alpha/2) \end{bmatrix} \quad (14)$$

The transformation starts with the angle $\alpha = 0$ and $\mathbf{D}_0^{Pitsch}(0)$ is the identity matrix. After the distortion, the $[\bar{1}10]_\gamma$ axis is rotated by an angle of $\pi/2 - \arccos(1/3) = \arcsin(1/3)$. Thus, the fcc-bcc transformation is completed when the rotation angle reaches $\alpha_{max} = \arcsin(1/3)$. By using equation (13) with $\sin(\alpha_{max}) = 1/3$ and $\cos(\alpha_{max}) = \sqrt{8}/3$, the distortion matrix of the complete Pitsch transformation becomes:

$$\mathbf{D}_0^{Pitsch} = \begin{bmatrix} \dfrac{1+\sqrt{2}}{3} & \dfrac{2-\sqrt{2}}{3} & 0 \\ \dfrac{1-\sqrt{2}}{3} & \dfrac{2+\sqrt{2}}{3} & 0 \\ 0 & 0 & 2/\sqrt{3} \end{bmatrix} \quad (15)$$

This is the matrix already reported in ref. [18]. It can be diagonalized and the eigenvalues are $(1, \sqrt{8}/3, 2/\sqrt{3}) \approx (1, 0.943, 1.155)$. One can also apply a polar decomposition to this matrix and its intermediate state matrices given in equation (13), such that $\mathbf{D}_0^{Pitsch}(\alpha) = \mathbf{R}^{Pitsch}(\alpha).\mathbf{S}^{Pitsch}(\alpha)$ where $\mathbf{R}^{Pitsch}$ is a (rigid body) rotation matrix and $\mathbf{S}^{Pitsch}$ a symmetric matrix given by

$$\mathbf{S}^{Pitsch}(\alpha) = \sqrt{{}^T\mathbf{D}_0^{Pitsch}(\alpha)\mathbf{D}_0^{Pitsch}(\alpha)} \tag{16}$$

The symbol $^T$ at the left of a matrix or a vector means "transpose". The equation becomes after calculations and comparison with equation (8)

$$\mathbf{S}^{Pitsch}(\alpha) = \mathbf{D}_0^{Bain}(\alpha) \tag{17}$$

Therefore, the matrix takes the form $\mathbf{D}_0^{Pitsch} = \mathbf{R}^{Pitsch}.\mathbf{D}_0^{Bain}$. The rotation matrix $\mathbf{R}^{Pitsch}$ of this decomposition appears naturally in the present approach whereas it is "artificially" created in the PTMC in order to respect the "invariant line criterion". When the transformation is completed ($\alpha = \alpha_{max}$), $\mathbf{R}^{Pitsch}$ is a rotation of 9.73° around $[00\bar{1}]_\gamma$.

## 5. Model of atomic displacements with Kurdjumov-Sachs OR

The KS OR is $[110]_\gamma // [111]_\alpha$, $[1\bar{1}2]_\gamma // [11\bar{2}]_\alpha$ and $[\bar{1}11]_\gamma // [\bar{1}10]_\alpha$. These γ axes, after normalization, form the common orthonormal reference basis $\mathbf{B}_c$. We propose now a distortion that a) respects the hard-sphere packing of the iron atoms such that the atoms moves relatively to each other exactly as described previously for the final Bain or Pitsch OR, b) imposes that the close packed direction $[110]_\gamma$ remains invariant, and c) imposes that the close packed plane $(\bar{1}11)_\gamma$ remains unrotated. This approach can be considered as a smooth and continuous combination of the initial Kurdjumov-Sachs-Nishiyama (KSN) model with a hard-sphere condition such as in the Burgers and Bogers model. The calculations are quite tedious and may be skipped at the first reading to go directly to equation (31).

The distortion matrix can be calculated by choosing a primitive non-orthogonal frame $\mathbf{B}_p$ constituted by the normalized axes $\mathbf{x} = (1/\sqrt{2})[110]_\gamma$, $\mathbf{y} = (1/\sqrt{2})[101]_\gamma$ and $\mathbf{z} = [100]_\gamma$. The $[110]_\gamma$ and $[101]_\gamma$ directions define the $(\bar{1}11)_\gamma$ plane that is transformed into the $(\bar{1}10)_\alpha$ plane by the distortion. By considering the Bain matrix given in equation (6), the angle β between the $[110]_\gamma$ and $[101]_\gamma$ axes appears as a function of the angle α:

$$\cos(\beta) = \sin^2(\pi/4 + \alpha/2) \tag{18}$$

For the fcc phase $\alpha = 0$ and $\beta = 60°$. When the transformation is complete, $\alpha = \alpha_{max} = \arcsin(1/3) = 19.4°$, and β can be calculated using the equation $\sin^2(\pi/4 + \alpha/2) = \frac{1-\sin\alpha}{2}$. It follows that β is such that $\cos(\beta) = 1/3$, i.e. $\beta = 70.5°$, which is the angle between the $[111]_\alpha$ and $[11\bar{1}]_\alpha$ directions of the $(\bar{1}10)_\alpha$ plane of the bcc phase.

Now, the effect of the distortion on the axis $\mathbf{z} = [100]_\gamma$ is calculated such that the hard-sphere packing is continuously respected. The geometrical view of the problem is illustrated in Fig. 6. Initially, in the fcc phase the three iron atoms in points O, P, K are in contact and form a triangle belonging to the $(\bar{1}11)_\gamma$ plane. The iron atom in M, at the head of the arrow **PM** =

[100]$_\gamma$, is in the upper layer 1/3.($\bar{1}$11)$_\gamma$ plane of the fcc stacking sequence ABCABC. As the angle β = $\widehat{OPK}$ = 60° is progressively increased to 70.5°, the atom in M is moved such that the distance PM is shorten while the distance PO = PK = MO = MK remains constant and equal to the atom diameter, i.e. $\sqrt{2}/2$. Thus, the projection H of M on the ($\bar{1}$11)$_\gamma$ plane remains along on the line **x+y**, initially [211]$_\gamma$. By considering the Bain matrix given in equation (6), the angle γ between the [211]$_\gamma$ and [100]$_\gamma$ directions appears as a function of the angle α:

$$\cos(\gamma) = \frac{\sqrt{2}\sin(\pi/4 - \alpha/2)}{\sqrt{1 + \sin^2(\pi/4 - \alpha/2)}} \quad (19)$$

and thus

$$\sin(\gamma) = \sqrt{\frac{1 - \sin^2(\pi/4 - \alpha/2)}{1 + \sin^2(\pi/4 - \alpha/2)}} \quad (20)$$

Now, let us associate the orthonormal basis **B**$_s$ = (**x**$_s$, **y**$_s$, **z**$_s$) to the basis **B**$_p$ = (**x**, **y**, **z**) as usually done for the structural tensor, i.e. **x**$_s$ is parallel to **x**, **y**$_s$ belongs to the (**x**, **y**) plane, and **z**$_s$ is oriented in the same direction than **z** relatively to the (**x**, **y**) plane. Since the **x** direction remains constant and the (**x**, **y**) plane is unrotated during the transformation, the basis **B**$_s$ remains unchanged despite the distortion of the basis **B**$_p$. The coordinates of the **x**, **y** and **z** vectors in the basis **B**$_s$ gives the coordinate transfomation matrix from **B**$_s$ to **B**$_p$, which is a function of the angles β and γ

$$[\mathbf{B}_s \to \mathbf{B}_p(\beta,\gamma)] = \begin{bmatrix} 1 & \cos(\beta) & \|z\|\cos(\gamma)\cos(\beta/2) \\ 0 & \sin(\beta) & \|z\|\cos(\gamma)\sin(\beta/2) \\ 0 & 0 & \|z\|\sin(\gamma) \end{bmatrix} \quad (21)$$

The norm of the vector **z** = [100]$_\gamma$ is given by equation (6): $\|z\| = \sqrt{2}\sin(\pi/4 - \alpha/2)$.

By using equations (18) to (20) and trigonometric rules, it can also be checked that

$$\begin{cases} \cos(\beta/2) = \sqrt{\dfrac{1 + \sin^2(\pi/4 - \alpha/2)}{2}} \\ \sin(\beta/2) = \sqrt{\dfrac{1 - \sin^2(\pi/4 - \alpha/2)}{2}} \end{cases} \quad (22)$$

Thus, equation (21) can be written

$$[\mathbf{B}_s \to \mathbf{B}_p(\beta, \alpha)] = \begin{bmatrix} 1 & \cos(\beta) & \sqrt{2}\sin^2(\pi/4 - \alpha/2) \\ 0 & \sin(\beta) & \sqrt{2}\sin^2(\pi/4 - \alpha/2)\sqrt{\dfrac{1 - \sin^2(\pi/4 - \alpha/2)}{1 + \sin^2(\pi/4 - \alpha/2)}} \\ 0 & 0 & \sqrt{2}\sin(\pi/4 - \alpha/2)\sqrt{\dfrac{1 - \sin^2(\pi/4 - \alpha/2)}{1 + \sin^2(\pi/4 - \alpha/2)}} \end{bmatrix} \quad (23)$$

By using equation (18) and noting $X = \cos(\beta)$, equation (23) can be written

$$[\mathbf{B}_s \to \mathbf{B}_p(\beta)] = \begin{bmatrix} 1 & X & \sqrt{2}X \\ 0 & \sqrt{1 - X^2} & \sqrt{2}X\sqrt{\dfrac{1 - X}{1 + X}} \\ 0 & 0 & \sqrt{2}\sqrt{X}\sqrt{\dfrac{1 - X}{1 + X}} \end{bmatrix} \text{ with } X = \cos(\beta) \quad (24)$$

The initial state of the fcc phase is defined by $\beta = 60°$, $X = 1/2$ and thus

$$[\mathbf{B}_s \to \mathbf{B}_p(60°)] = \begin{bmatrix} 1 & 1/2 & \sqrt{2}/2 \\ 0 & \sqrt{3}/2 & 1/\sqrt{6} \\ 0 & 0 & 1/\sqrt{3} \end{bmatrix} \quad (25)$$

Its inverse is

$$[\mathbf{B}_p(60°) \to \mathbf{B}_s] = \begin{bmatrix} 1 & -1/\sqrt{3} & -\sqrt{2/3} \\ 0 & 2/\sqrt{3} & -\sqrt{2/3} \\ 0 & 0 & \sqrt{3} \end{bmatrix} \quad (26)$$

During the fcc-bcc transformation, X varies from 1/2 ($\beta = 60°$) to 1/3 ($\beta = 70.5°$).

The angular distortion matrix calculated from the initial fcc phase to an intermediate state parameterized by the angle $\beta$ is given by the matrix

$$\mathbf{D}_p^{KS}(\beta) = [\mathbf{B}_p(60°) \to \mathbf{B}_p(\beta)] = [\mathbf{B}_p(60°) \to \mathbf{B}_s][\mathbf{B}_s \to \mathbf{B}_p(\beta)] \quad (27)$$

$$= \begin{bmatrix} 1 & X - \sqrt{\dfrac{1 - X^2}{3}} & \dfrac{1}{\sqrt{3}}\sqrt{\dfrac{1 - X}{1 + X}}(\sqrt{6}X\sqrt{\dfrac{1 + X}{1 - X}} - \sqrt{2}X - 2\sqrt{X}) \\ 0 & 2\sqrt{\dfrac{1 - X^2}{3}} & \dfrac{2}{\sqrt{3}}\sqrt{\dfrac{1 - X}{1 + X}}(\sqrt{2}X - \sqrt{X}) \\ 0 & 0 & \sqrt{6}\sqrt{X}\sqrt{\dfrac{1 - X}{1 + X}} \end{bmatrix} \text{ with } X = \cos(\beta)$$

This matrix gives the distortions along the axes **x**, **y**, **z** of $\mathbf{B}_p$. One can associate to this basis the coordinate transformation matrix $[\mathbf{B}_0 \to \mathbf{B}_p(60°)]$ given by

$$[\mathbf{B}_0 \to \mathbf{B}_p(60°)] = \begin{bmatrix} 1/\sqrt{2} & 1/\sqrt{2} & 1 \\ 1/\sqrt{2} & 0 & 0 \\ 0 & 1/\sqrt{2} & 0 \end{bmatrix} \quad (28)$$

Its inverse is

$$[\mathbf{B}_p(60°) \to \mathbf{B}_0] = \begin{bmatrix} 0 & \sqrt{2} & 0 \\ 0 & 0 & \sqrt{2} \\ 1 & -1 & -1 \end{bmatrix} \quad (29)$$

These matrices allow us to calculate the distortion matrix in the basis $\mathbf{B}_0$

$$\mathbf{D}_0^{KS}(\beta) = [\mathbf{B}_0 \to \mathbf{B}_p(60°)] \, \mathbf{D}_p^{KS}(\beta) \, [\mathbf{B}_p(60°) \to \mathbf{B}_0] = d_{ij}^{KS}(\beta) \quad (30)$$

with the components $d_{ij}^{KS}(\beta)$ in line $i$ and column $j$ given by

$$\begin{cases} d_{11}^{KS}(\beta) = \frac{1}{\sqrt{6}} \sqrt{\frac{1-X}{1+X}} \left(\sqrt{2}X + 2\sqrt{X}\right) + X \\ d_{21}^{KS}(\beta) = -\frac{1}{\sqrt{6}} \sqrt{\frac{1-X}{1+X}} \left(\sqrt{2}X + 2\sqrt{X}\right) + X \\ d_{31}^{KS}(\beta) = \frac{2}{\sqrt{6}} \sqrt{\frac{1-X}{1+X}} \left(\sqrt{2}X - \sqrt{X}\right) \\ d_{12}^{KS}(\beta) = 1 - X - \frac{1}{\sqrt{6}} \sqrt{\frac{1-X}{1+X}} \left(\sqrt{2}X + 2\sqrt{X}\right) \\ d_{22}^{KS}(\beta) = 1 - X + \frac{1}{\sqrt{6}} \sqrt{\frac{1-X}{1+X}} \left(\sqrt{2}X + 2\sqrt{X}\right) \\ d_{32}^{KS}(\beta) = -\frac{2}{\sqrt{6}} \sqrt{\frac{1-X}{1+X}} \left(\sqrt{2}X - \sqrt{X}\right) \\ d_{13}^{KS}(\beta) = \sqrt{\frac{1-X^2}{3}} - \frac{1}{\sqrt{6}} \sqrt{\frac{1-X}{1+X}} \left(\sqrt{2}X + 2\sqrt{X}\right) \\ d_{23}^{KS}(\beta) = -\sqrt{\frac{1-X^2}{3}} + \frac{1}{\sqrt{6}} \sqrt{\frac{1-X}{1+X}} \left(\sqrt{2}X + 2\sqrt{X}\right) \\ d_{33}^{KS}(\beta) = 2\sqrt{\frac{1-X^2}{3}} - \frac{2}{\sqrt{6}} \sqrt{\frac{1-X}{1+X}} \left(\sqrt{2}X - \sqrt{X}\right) \end{cases} \quad \text{with } X = \cos(\beta) \quad (31)$$

This matrix gives all the intermediate states resulting from the continuous distortion of the fcc lattice ($\beta = 60°$, $X = 1/2$) toward the final distorted fcc lattice ($\beta = 70.5°$, $X = 1/3$), which is actually the bcc lattice in KS OR. The complete distortion is obtained with $X=1/3$:

$$\mathbf{D}_0^{KS} = \begin{bmatrix} \sqrt{6}/18 + 2/3 & -\sqrt{6}/18 + 1/3 & \sqrt{6}/6 - 1/3 \\ -\sqrt{6}/18 & \sqrt{6}/18 + 1 & -\sqrt{6}/6 + 1/3 \\ \sqrt{6}/9 - 1/3 & -\sqrt{6}/9 + 1/3 & \sqrt{6}/3 + 1/3 \end{bmatrix} \quad (32)$$

This matrix has 1, 1 and 1.088 for eigenvalues. The eigenvector associated to 1.008 is a vector close to $[4, 7, -3]_\gamma$. The linear subspace associated to 1 (multiplicity =2) is reduced to the line along the eigenvector $[110]_\gamma$. This implies that the matrix $\mathbf{D}_0^{KS}$ can't be diagonalized. However, it can be indirectly diagonalized by polar decomposition: $\mathbf{D}_0^{KS} = \mathbf{R}^{KS} \mathbf{S}^{KS}$ with $\mathbf{R}^{KS}$ a rotation matrix and $\mathbf{S}^{KS}$ a symmetric matrix given by

$$\mathbf{S}^{KS} = \sqrt{{}^T\mathbf{D}_0^{KS} \mathbf{D}_0^{KS}} \quad (33)$$

which becomes after calculations, and comparison with equation (8)

$$\mathbf{S}^{KS} = \mathbf{D}_0^{Bain} \quad (34)$$

Therefore, as for Pitsch, the matrix of distortion associated to the KS OR takes the form $\mathbf{D}_0^{KS} = \mathbf{R}^{KS}\mathbf{D}_0^{Bain}$. The rotation $\mathbf{R}^{KS}$ is a rotation of 11.06° around $[21, 25, -60]_\gamma$.

## 6. Comparison of the distortions associated to Bain, Pitsch and KS orientations

### 6.1. 3D representations

The distortions calculated for Bain, Pitsch and KS orientations are represented in 3D in Fig. 7 by using VPython as computing language. The initial fcc cube (8x8 cells) is coloured in blue, with the **X**, **Y** and **Z** directions of the cube oriented along the $[100]_\gamma$, $[010]_\gamma$ and $[001]_\gamma$ directions. With Bain, the crystal is compressed along **X** and dilated along **Y** and **Z**, and these directions are unrotated. With Pitsch, the direction **Z** is unrotated and the direction **X+Y** is along $[110]_\gamma$ (noted **N** for neutral) is invariant. With KS, the **X**, **Y** and **Z** directions are rotated but the direction **X+Y** is invariant and the $(\bar{1}11)_\gamma$ plane is unrotated. The fcc crystals after distortions are coloured in red. It can be checked visually that they are effectively bcc crystals. The movies of the fcc-bcc continuous distortions for the Bain, NW and KS ORs are in the Supplementary Materials.

### 6.2. Correspondence matrices for the directions and planes

The matrices $\mathbf{D}_0^{OR}$ of equations (8), (15) and (32) give, in the initial fcc basis $\mathbf{B}_0^\gamma$, the images of the direction $\mathbf{u}_0^\gamma$ by the distortions leading to the Bain, Pitsch and KS ORs, respectively. The inverse of their transpose gives the distortion of the planes $\mathbf{g}_0^\gamma$. The images are

$$\mathbf{u}_0'^\gamma = \mathbf{D}_0^{OR}\mathbf{u}_0^\gamma \text{ and } \mathbf{g}_0'^\gamma = {}^T\left(\mathbf{D}_0^{OR}\right)^{-1}\mathbf{g}_0^\gamma, \text{ for OR = Bain, Pitsch or KS.} \quad (35)$$

It is possible to calculate these images in the final bcc reference frame by using the coordinate transformation matrices, which are calculated from the orientation relationships.

The OR can be expressed by a triplet of pairs of direction normalized vectors which define a common basis $\mathbf{B}_c^\gamma / a_\gamma = (\mathbf{x}_c^\gamma, \mathbf{y}_c^\gamma, \mathbf{z}_c^\gamma) / a_\gamma = \mathbf{B}_c^\alpha / a_\alpha = (\mathbf{x}_c^\alpha, \mathbf{y}_c^\alpha, \mathbf{z}_c^\alpha) / a_\alpha$. The (α→γ) coordinate transformation matrix $\mathbf{T}_{\alpha\gamma}$ for the directions is then given by

$$\mathbf{T}_{\alpha\gamma} = \left[\mathbf{B}_0^\alpha \to \mathbf{B}_0^\gamma\right] = \left[\mathbf{B}_0^\alpha \to \mathbf{B}_c^\gamma\right]\left[\mathbf{B}_0^\gamma \to \mathbf{B}_c^\gamma\right]^{-1} = \frac{a_\gamma}{a_\alpha}\left[\mathbf{B}_0^\alpha \to \mathbf{B}_c^\alpha\right]\left[\mathbf{B}_0^\gamma \to \mathbf{B}_c^\gamma\right]^{-1} = \sqrt{\frac{3}{2}}\mathbf{R}_{\alpha\gamma} \quad (36)$$

where $\mathbf{R}_{\alpha\gamma} = \sqrt{\frac{2}{3}}\mathbf{T}_{\alpha\gamma}$ is a rotation matrix between the fcc and bcc cubic crystals. The coordinate transformation matrix for the planes are the inverse of the transpose of $\mathbf{T}_{\alpha\gamma}$ which is equal to $\sqrt{2/3}\mathbf{R}_{\alpha\gamma} = 2/3\,\mathbf{T}_{\alpha\gamma}$. The distorted vectors of directions and planes in the final α bcc reference frame $\mathbf{B}_0^\alpha$ are given by

$$\mathbf{u}'^\alpha_0 = \mathbf{T}_{\alpha\gamma}\mathbf{u}'^\gamma_0 \text{ and } \mathbf{g}'^\alpha_0 = 2/3\,\mathbf{T}_{\alpha\gamma}\mathbf{g}'^\gamma_0 \qquad (37)$$

Combining equations (35) and (37) gives the correspondence matrices

$$\mathbf{C}_{\alpha\gamma} = \mathbf{T}_{\alpha\gamma}\mathbf{D}_0^{OR} \text{ and } \mathbf{C}^*_{\alpha\gamma} = 2/3\,\mathbf{T}_{\alpha\gamma}^{\,T}\left(\mathbf{D}_0^{OR}\right)^{-1} \qquad (38)$$

which permit to calculate directly the images of the directions and planes in the final bcc reference frame:

$$\mathbf{u}'^\alpha_0 = \mathbf{C}_{\alpha\gamma}\mathbf{u}^\gamma_0 \text{ and } \mathbf{g}'^\alpha_0 = \mathbf{C}^*_{\alpha\gamma}\mathbf{g}^\gamma_0 \qquad (39)$$

The distortion matrices, the rotational part $\mathbf{R}_{\alpha\gamma}$ associated to the coordinate transformation matrices and the images of some low-index direct and reciprocal directions noted in the final α bcc basis are given for Bain, Pitsch and KS ORs in Table 1. It can be checked that they are the same whatever the distortion matrix, Bain, Pitsch or KS of equations (8), (15) and (32). This results from the fact that the correspondence matrices are of Bain type, i.e. with vector components of type <100>, <110> and <$\bar{1}$10>. The calculations show that the correspondence matrices of Bain, Pitsch and KS are equal if one accepts the equivalence due to the three-fold (**x**→**y**→**z**) symmetry. This comes from the fact that the OR of the bcc phase has been chosen in order that the $[110]_\gamma$ axis is transformed into the $[111]_\alpha$ axis without imposing a coherence on the notation for the other axes.

**7. Discussions**

**7.1. Difference of the iron atom size in the fcc and bcc phases**

The continuous distortion matrices associated to Bain, Pitsch and KS ORs have been calculated by assuming that the size of the iron atoms remains constant during the transformation, which is equivalent to equation (5) for the lattice parameters. Actually, it is possible to introduce the difference of atom size by multiplying the distortion matrices by the real factor *k*, ratio of iron atom radius in the bcc phase $r_\alpha$ to its radius in the fcc phase $r_\gamma$, i.e. $k = r_\alpha/r_\gamma = \dfrac{\sqrt{3}a_\alpha}{\sqrt{2}a_\gamma}$. The k ratio is generally around 0.96 or 0.97. The 3 or 4% of difference is not negligible if it is compared to the 8% of difference of volume change, but this difference is acceptable for a first qualitative model because it is well known that hydrostatic part of the strains does not play a primary role on the transformation in comparison to the shear components [11]. To be more quantitative one would have to determine how *k* varies from the initial fcc phase to the final bcc phase along the transformation path during the few picoseconds of its completion. $k$ should be a real function of the angular distortion parameter which, for Bain and Pitsch ORs is equal to the angle α in the range from 0 to arcsin(1/3), and for KS OR is equal to the angle β in the range from arcos(1/2) to arcos(1/3). Its extreme values are $k = 1$ and $k = r_\alpha/r_\gamma$ for complete transformation. Different simple models can be proposed such as an abrupt change at the start of the transformation, or an abrupt change at the end of the transformation, or a smooth linear change during the transformation. Determining the exact nature of $k$ as function of the distortion angle would require a deep physical investigation of the effect of the coordination on the electronic and magnetic properties of the iron atoms; which is beyond the scope of the paper.

**7.2. Proposition of a new transition order parameter**

To our knowledge, up to now, only complex order parameters have been attributed to fcc-bcc martensitic transformations, most of them associated to the stress field around the martensite [20]. More generally, as frankly said by Clapp in 1995 [21] "*Certainly one of the stumbling blocks in defining a martensitic transformation is that there is no obvious order parameter associated with it, such as one has with ferromagnetic transformations (magnetic moment), order-disorder (long range order parameter), etc*". In our approach, whatever the matrix, associated to Bain, Pitsch or KS OR, assuming that the distortion respects the hard sphere packing allows us to introduce a unique natural order parameter, which is the angular parameter of the distortion, i.e. α in equations (6) and (14) for Bain and Pitsch, respectively, and β or $X = \cos(\beta)$ in equation (31) for KS. This parameter could probably be used in phase field approaches.

**7.3. What is the natural fcc-bcc distortion mechanism?**

In the paper, we call "natural" the hypothetical distortion of a very small free fcc single-crystal, isolated from any external stress field, that transforms into one variant of bcc martensite. What could be the natural fcc-bcc martensite distortion? Does it follow the distortion matrix (8), (15) or (32), corresponding to Bain, Pitsch or KS, respectively? A pure Bain distortion seems possible. Indeed, it is argued in the PTMC that the Bain OR is not observed experimentally because of the strain accommodation of the surrounding austenite matrix. It means that the rigid body rotation would be a consequence of the transformation and not an intrinsic part of the mechanism. However, we don't find any crystallographic argument that could support the assumption that the distortion naturally respects a 4-fold $<100>_\gamma$ axis of austenite (the compression axis). If one considers only the strains associated to the distortion, there is no difference between the three paths presented in Fig. 7 because all are composed of the same strains and differ only by the rotation associated to the strain. Actually, Pitsch OR, and thus equation (15), seems more probable than Bain, because Pitsch

OR was experimentally observed in thin TEM lamellas of Fe-N alloys after quenching [19], i.e. in samples nearly free of any surrounding austenite. By using distortion matrices such as Pitsch, the rigid body rotation becomes an intrinsic component of the distortion, and the Bain distortion, as expressed by the matrix (8), is just another component of the distortion which characterizes the hard sphere packing rule. Is Pitsch the natural distortion mechanisms for all the martensitic alloys as assumed in ref. [18]? Actually, KS is the most often reported OR in bulk steels and KS was also observed by Maki and Wayman in TEM lamella of Fe-Ni-C [22]. Therefore, it is probable that Pitsch only applies to Fe-N alloys whereas KS applies to the other steels. Now, we would be inclined to think that KS is probably the right natural transformation (see next section). However, it just an assumption that should have to be confirmed by experimental observations. For example, one could select by EBSD an grain of metastable austenite and oriented such that the sample axes **X**, **Y**, **Z** are along the $<100>_\gamma$ directions as in Fig. 7, cut a small cube by Focus Ion Beam (FIB) and maintain it at the tip of a very thin copper needle on which an absolute reference frame has been carved by FIB. By cooling the sample below $M_s$, the cube should be transformed into a parallelepiped, as illustrated in Fig. 8. If Bain is respected, one should not observe any rotation of the sample axes **X**, **Y**, **Z**; if Pitsch is respected only **Z** should remain unrotated, and if KS is respected all the three axes should be rotated. Other experiments are also possible but will not be detailed here.

**7.4. What is the fcc-bcc matrix-embedded distortion?**

As mentioned previously, considering the strains associated to the distortions matrices (8), (15) or (32) does not permit to differentiate them. However, if one considers the atomic displacements, which seems more adequate to take into consideration the surrounding matrix, the paths are not equivalent anymore. Indeed, it is possible to calculate the mean displacement of the atoms in a cell from the distortion matrices (8), (15) or (32). The matrices **D-I** are applied to the 14 atoms of the cells, with atoms at the corners counting for ¼ and those in the middle of the faces for ½, and the norms are arithmetically averaged. For an initial fcc matrix of $a_\gamma$ = 3.57 Å, the mean displacement is 0.591 Å for Bain, 0.476 Å for Pitsch, and 0.520 Å for KS. Therefore, Pitsch has the lowest mean displacement. Thus, here again, one could tempted to admit Pitsch as the most appropriate fcc-bcc matrix-embedded distortion in bulk steels. However, we came to believe that KS could be more correct because many results reported in literature, including our recent automated treatments of EBSD maps [23], showed that even if the martensitic grains exhibit spreading between all the classical ORs, Pitsch is generally found in low proportion in comparison with KS, and the central part of the laths along the long lath direction is not in Pitsch but in KS OR. Moreover, many TEM results have shown that the isolated martensite laths or plates are in KS OR with the surrounding fcc matrix. This experimental fact can't be explained with Bain or Pitsch with symmetry arguments. Indeed, the parent fcc matrix and the daughter bcc crystal share some common symmetries that form a subgroup of the parent fcc point group, called intersection group. The shape of the bcc crystal with the orientation of its faces should respect this intersection group and the number of variants is the order of the parent group divided by the order of this intersection group [24][25]. The intersection group is composed of four elements for Bain and Pitsch and only two elements for KS (identity and inversion), that is why there are 12 variants for the former ORs and 24 for the latter. It is difficult to explain why 24 variants are always formed if the natural transformation is Bain or Pitsch, even by taking into account the relaxation modes of the fcc matrix (dislocations, twinning etc) because they are symmetrically equivalent in the intersection group and should all be activated without changing the shape symmetry and variant number. The PTMC introduces a dissymmetry in these relaxation modes by choosing arbitrarily one of them, but such a choice signifies that the transformation itself depends on an initial asymmetry inside the fcc matrix (pre-existing dislocations, slip bands etc.). That possibility can't be excluded but we think simpler to assume from the beginning that the natural distortion is the one that leads to the KS OR without requiring a hypothetical external symmetry breaking induced by defects. In the rest of the paper, we will assume that KS is the natural distortion of the fcc-bcc martensitic transformation.

**7.5. Particularity of the distortion associated to the KS OR**

For the matrices $\mathbf{D}_0^{OR}$ calculated with Bain, Pitsch or KS OR, it can be checked that its determinant, i.e. the volume change of the transformation, is $\det(\mathbf{D}_0^{OR}) = \frac{4}{3}\sqrt{2/3} \approx 1.088$, as expected for a transformation between bcc and fcc phases linked by a hard sphere model. This point can also be proved from the polar decomposition of the matrices. The volume change results from the lattice dilatation along the eigenvectors of the matrices (8) and (15) associated to Bain and Pitsch ORs; these directions exist because the matrices can be diagonalized. The particularity of the distortion matrix (32) associated to the KS OR is that it can't be diagonalized. It has 1, 1 and 1.088 as characteristic roots, but it is not of shear nor invariant-plane strain type. Let us recall here that an invariant-plane strain $\mathbf{P}$ can be written $\mathbf{P} = \mathbf{I} + \mathbf{d}.^T\mathbf{p}$, where $\mathbf{I}$ is the identity matrix, $\mathbf{d}$ is the displacement (column) vector and $^T\mathbf{p}$ is the (line) vector normal to the invariant plane [8]. Its characteristic roots are 1, 1 and $1 + ^T\mathbf{p}.\mathbf{d}$. The two unit values are correlated to the invariant plane and the volume change given by the dilatation component $\delta = ^T\mathbf{p}.\mathbf{d}$ occurs perpendicularly to the invariant plane. Similarly to the invariant plane strain $\mathbf{P}$, the volume change of $\mathbf{D}_0^{KS}$ occurs by the dilatation along one unique axis, which is the eigenvector associated to the value 1.088, i.e. here the vector $[4,7,-3]_\gamma$, but contrarily to **P** this vector belongs to the unrotated $(\bar{1}11)_\gamma$ plane, which means that the totality of the volume change is generated by the angular distortion inside this plane. This can be also understood by calculating the surface change associated to the transformation from the triangle on the $(\bar{1}11)_\gamma$ plane formed by the $[110]_\gamma$ and $[101]_\gamma$ diagonals (with angle of 60°) to the distorted triangle on the $(\bar{1}10)_\alpha$ plane formed by the $[111]_\gamma$ and $[11\bar{1}]_\gamma$ diagonals (with angle of 70.5°), as illustrated in Fig. 9. It is $S_{bcc}/S_{fcc} = \sin(70.5°)/\sin(60°) \approx 1.088$. The plane of distortion $(\bar{1}11)_\gamma$ is unrotated, and the directions it contains are not dilated or contracted but their angle changes. The totality of the volume change comes from this angular distortion.

The fact that the volume change of the complete transformation remains localized on the $(\bar{1}11)_\gamma$ plane also implies that the atom in M located in the upper layer $1/3.(\bar{1}11)_\gamma$ has moved such that the distances MH on Fig. 6 at the transformation start (X=1/2) and at the transformation end (X=1/3) are equal. However, it is worth pointing out that the displacement of the atom M is not of shear type because the distance MH is not constant during the transformation. Indeed, MH is given by the scalar product between the direction $[\bar{1}11]_\gamma$ and the image of the direction $[100]_\gamma$ by the deformation matrix (31), it is:

$$MH = \sqrt{\frac{2X(1-X)}{1+X}} \tag{40}$$

Another way to verify that MH is not constant is to consider the surface change of the $(\bar{1}11)_\gamma$ plane and the volume change of the lattice; they do not obey the same law. Indeed, the surface change S(X) is a function of $\sin(\beta) = \sqrt{1-X^2}$, and the volume change V(X) is a function of $\sqrt{X}(1-X)$, as deduced from the matrix (27). Equation (40) results from the formula V(X) = 3.MH.S(X). The lateral displacement of the atom in M is possible despite its uncompressible contact with the atoms in O and P thanks to the angular distortion β which increases the distance OP as the atom M moves (the atoms in O and P loose contact when the transformation starts).

The distortion matrix (32) associated to the KS OR has the properties of an invariant line strain (ILS) and thus could constitute an interesting alternative to the **RB** term used in the PTMC. The general mathematics associated to angular distortions is probably worth being developed. The concept of disclinations (special pile-ups of dislocations that create rotational discontinuities of lattices) seems very relevant to accommodate the 60°→70.5° angular distortion on the $(\bar{1}11)_\gamma$ plane, as suggested in the one-step model (supplementary materials 4 of [18]). We also recall that the matrix (32) has by construction the unique property of letting invariant both a close-packed direction and a close-packed plane. Indeed, $[110]_\gamma$ is invariant by $\mathbf{D}_0^{KS}$ and is transformed into $[111]_\alpha$ by $[\mathbf{B}_0^\alpha \to \mathbf{B}_0^\gamma]$, and $(\bar{1}11)_\gamma$ is invariant by $^T(\mathbf{D}_0^{KS})^{-1}$ and is transformed into $(110)_\alpha$ by $\frac{2}{3}[\mathbf{B}_0^\alpha \to \mathbf{B}_0^\gamma]$.

**7.6. The one-step model with KS OR**

We have shown that it is possible to obtain both the good lattice and the correct orientation relationship with a unique matrix that respect the hard sphere packing of the atoms and that can be expressed by polar decomposition as a product of a rotation and a Bain distortion. Such a distortion is not an invariant plane strain; so, in order to try to predict the habit planes as in PTMC, it could be possible to combine it with lattice invariant shear such as twins, slip, or coupling the variants by pairs, as recently detailed in [26]. In bcc martensite of low carbon steels formed at high temperature, the accommodation of the fcc/bcc lattice incompatibilities is realized by special configurations of dislocations. To our point of view, these configurations (disclinations) are at the origin of the OR spreading observed in the EBSD or X-ray diffraction pole: there is a unique distortion and a unique local OR (KS) and the OR spreading comes from the accommodating strain field generated by the transformation itself [17]. This idea is at the core of the one-step model [18]. That model was applied with Pitsch but it can also be applied with KS; in that case, the sequence of events should be slightly modified in comparison to ref. [18]. One has to imagine that the nucleation of martensite in a surrounding austenite field occurs by a distortion leading to KS by equation (32), and that the Pitsch and NW orientations result from the growth of the martensite in the austenite deformed by the continuous **A** and **B** rotations that were experimentally observed in pole figures [17][18]. The modified scheme of this scenario is proposed in Fig. 10. Using KS instead of Pitsch as unique distortion matrix seems more attractive for the reasons mentioned in the previous section. Moreover, the rotation **B** with Pitsch is combined with rotation **A** (see section 4.2 of [18]), whereas it is pure with KS because it directly results from the angular $(\bar{1}11)_\gamma \to (110)_\alpha$ distortion as explained in section 7.5. We are currently working to find an exact mathematical method to extract the two experimentally observed rotations **A** and **B** directly from the distortion matrix. The first analysis is very encouraging. It is for example striking to note that the rotation **A** corresponds to the angular distortion of the unrotated $(\bar{1}11)_\gamma$ plane, i.e. to the rotation of the direction $[101]_\gamma$ around $[\bar{1}11]_\gamma$, and that the rotation **B** corresponds to the angular distortion around the invariant $[110]_\gamma$ direction , i.e. to the rotation of the $(1\bar{1}1)_\gamma$ plane around $[110]_\gamma$; exactly as if rotations **A** and **B** were the angular distortive parts of the martensitic distortion in the direct and reciprocal space, respectively.

## 7.7. Prediction of the {225} habit planes

Because it seems possible that some bcc martensite are mostly accommodated by dislocations, we tried to check whether the distortion matrix $\mathbf{D}_0^{KS}$ could be directly used to predict some habit planes without requiring additional arguments such as multiple slips, twinning or variant pairing. Since the distortion matrix has no invariant plane, a weaker constraint was investigated. We realized that, even if not fully invariant, the habit plane should at least correspond to an unrotated plane because a rotated plane would create huge displacements at the interface with the fcc matrix far from the rotation axis. The crystallographic condition to obtain an unrotated plane can be simply written as follows. The distortion matrix for the directions is $\mathbf{D}_0^{KS}$, the distortion matrix for the planes is $\mathbf{D}_0^{KS*} = {}^T\!\left(\mathbf{D}_0^{KS}\right)^{-1}$, the displacement matrix for the directions is $\mathbf{F}_0^{KS} = \mathbf{D}_0^{KS} - \mathbf{I}$, and the displacement matrix for the planes is $\mathbf{F}_0^{KS*} = {}^T\!\left(\mathbf{D}_0^{KS}\right)^{-1} - \mathbf{I}$. The displacement of a plane $\mathbf{g}$, $\Delta\mathbf{g} = \mathbf{g}' - \mathbf{g} = \mathbf{F}_0^{KS*}\mathbf{g}$, can be decomposed into $\Delta\mathbf{g} = \Delta\mathbf{g}_{/\!/} + \Delta\mathbf{g}_\perp$ with $\Delta\mathbf{g}_{/\!/}$ parallel to $\mathbf{g}$ and $\Delta\mathbf{g}_\perp$ perpendicular to $\mathbf{g}$, as shown in Fig. 11. The part $\Delta\mathbf{g}_{/\!/}$ corresponds to the case where $\mathbf{g}'$ remain parallel to $\mathbf{g}$, i.e. to a modification of the interplanar distance without rotation. The part $\Delta\mathbf{g}_\perp$ corresponds to a rotation of the plane; it is equal to the rotation angle for small rotations. For any vector $\mathbf{g}$ isotropically oriented in space, i.e. oriented on the unit sphere $\|\mathbf{g}\| = 1$, one can calculate $\Delta\mathbf{g}$ and then the norm of $\Delta\mathbf{g}_\perp$ by

$$\|\Delta\mathbf{g}_\perp\| = \|\Delta\mathbf{g} - (\mathbf{g}.\Delta\mathbf{g}).\mathbf{g}\| = \|\mathbf{F}_0^{KS*}\mathbf{g} - (\mathbf{g}.\mathbf{F}_0^{KS*}\mathbf{g}).\mathbf{g}\| \tag{41}$$

The planes $\mathbf{g}$ unrotated by the distortion are those with $\|\Delta\mathbf{g}_\perp\| = 0$. This is actually one the simplest condition used in the O-lattice theory [27][14]. Analytical determination of the solutions of equation (41) is possible, but for this very preliminary work the calculations have been performed numerically with Mathematica. This also permits to get a better idea of the space of planar rotations. Indeed, for any $\mathbf{g}$ vector on the sphere, one can plot the value $\|\Delta\mathbf{g}_\perp\|$. The surface of $\mathbf{g}$ vectors is reported in the 3D space in Fig. 12a. Other representations showing the values $\|\Delta\mathbf{g}_\perp\|$ as function of the two spherical coordinate angles θ and φ are given in Fig. 12b and Fig. 12c. The surface exhibits four minima, two at θ ≤ π/2 and two at θ > π/2. The solutions at θ > π/2 are actually equivalent to those obtained at θ ≤ π/2 because of the centrosymmetric equivalence of lines (θ, φ) ≡ (π-θ, φ+π). The numerical values of the two minima located in the region θ ≤ π/2 and transformed back into Cartesian coordinates are $\mathbf{g}_1$ = (-0.5773, 0.5773, 0.5773) and $\mathbf{g}_2$ = (-0.3535, 0.3535, 0.8660). They are shown in Fig. 12d. The first solution $\mathbf{g}_1$ is the $(\bar{1}11)_\gamma$ plane, which was expected because this plane is unrotated by construction, but the second solution **g**$_2$ was unexpected, it is at only 0.5° of the well-known $(\bar{2}25)_\gamma$ plane that contains the common neutral line $[110]_\gamma = [111]_\alpha$ and orientated at 25° from the unrotated close-packed plane $(\bar{1}11)_\gamma = (\bar{1}10)_\alpha$. The $(\bar{2}25)_\gamma$ plane is transformed into the $(7, \bar{4}, \bar{3})_\alpha$ plane. The solution does not depend on the ratio $k = r_\alpha/r_\gamma = \dfrac{\sqrt{3}a_\alpha}{\sqrt{2}a_\gamma}$ and therefore is independent of the lattice parameters of the fcc and bcc phases. This result is very encouraging. It is worth recalling that the $\{225\}_\gamma$ habit planes have remained unexplained for a very long time and were "solved" with the PTMT in the 1960's by introducing double lattice invariant shears. A brief summary of the "saga" of the $\{225\}_\gamma$ planes can be found in [14]. A similar approach leading to the same results were already obtained by Jaswon and Wheeler in 1948 [28]; we discovered it only after writing the manuscript. This paper was considered as an important step at that time but was then largely forgotten due to the predominance of the PTMT.

In the present approach the $\{225\}_\gamma$ habit planes simply appear as planes unrotated by the fcc-bcc distortion. It is possible that the unrotated $(\bar{2}25)_\gamma$ plane is also made completely invariant by the accommodating configurations of dislocations at the origin of the continuous rotations **A** and **B**. It is also interesting to note that all the planes **g** between **g**$_1$ and **g**$_2$ and containing the neutral direction are at the valley of the surface with very low rotation values, which could explain why other habit planes such as $(\bar{5}57)_\gamma$ can be also formed.

We introduced a tetragonal distortion along the **a** axis in the calculations to see whether or not is could be possible to use similar criteria to "predict" the other families of habit planes such as the $\{2\,5\,9\}_\gamma$ and $\{3\,10\,15\}_\gamma$ habit planes, but the results are not yet satisfying, as already noticed by Jaswon and Wheeler [28]. For $c_\alpha/a_\alpha = 1.05$, the solution is close to $(\bar{5}\,3\,15)_\gamma$ which is at more than 10° from the $(\bar{5}\,2\,9)_\gamma$ or $(\overline{10}\,3\,15)_\gamma$ planes. The simplicity of the present approach has certainly reached its limit, and more complex calculations are required. Some approaches used in PTMC, such as twinning or variant pairing proved to be very efficient to model the $\{2\,5\,9\}_\gamma$ and $\{3\,10\,15\}_\gamma$ habit planes [9][13][14] and more recently the $\{5\,5\,7\}_\gamma$ habit planes [26]. More adapted and refined criteria such as those used in the O-lattice theory are also worth being investigated [14].

### 7.8. Perspectives

In the paper the continuous matrices associated to Bain, Pitsch and KS ORs have been determined. The method for calculating the distortion matrix for the NW OR would be analogous; the calculations are not reported here because we wanted to focus on Pitsch and KS ORs which are the only ORs that let invariant the close packed direction $[110]_\gamma = [111]_\alpha$.

Many works are required to check whether the present approach of martensitic transformation in steels and iron alloys is worth being continued. The natural distortion matrix should be confirmed by cooling small cube of metastable austenite shaped by FIB (section 7.3). Energetical calculations by molecular dynamics on systems composed of a fcc matrix with bcc martensite forming inside this matrix could also be bring new information, but would require further fundamental research on the nature of the accommodating configurations of dislocations and the possible link with the concept of disclination. It could also be interesting to investigate the possible interest of the angular distortive matrices to explain the tetragonal martensite morphologies, the variant selection mechanisms and the burst phenomena. For example it could be worth determining an energetic criterion of variant selection equivalent to the famous Patel and Cohen's one [29] but adapted to the matrix $\mathbf{D}_0^{KS}$ and its angular distortive character.

Since fcc-bcc martensite transformation, bcc-hcp martensite transformation and mechanical twinning share many common characteristics, the present approach seems to be generalizable to most of the metals with hard-sphere packing. Actually, the continuous angular distortive matrices of the fcc, bcc, hcp and fcc twinning transformations have already been calculated; the results will be detailed and discussed in a next paper.

## 8. Conclusions

The lattice distortions associated to the Bain, Pitsch and KS orientation relationships have been calculated with the simple assumption of a hard-sphere packing of the iron atoms. The continuous distortion matrices of the fcc-bcc transformations are functions of a unique parameter, the angle of planar distortion, which can constitute a reasonable transition order parameter. The matrix associated to the KS OR is very special; by construction it lets the close packed direction $[110]_\gamma$ invariant - transformed into $[111]_\alpha$- and the close packed plane $(\bar{1}11)_\gamma$ unrotated - transformed into $(\bar{1}10)_\alpha$-. The matrix has 1, 1 and 1.088 for eigenvalues, and, importantly, it is not of shear type. Indeed, the whole fcc/bcc volume change is associated to the angular distortion of the $(\bar{1}11)_\gamma$ unrotated plane. Beyond the calculations, the paper actually proposes a new paradigm for the fcc-bcc martensitic transformations. Up to now, these transformations were considered as "shear" transformations, nearly by definition: "*shear transformations are synonymous with martensitic transformation*" [13]. In our work, the fcc-bcc martensitic transformations appear as "angular distortive". This idea is actually not so far from the initial historical models of Kurdjumov and Sachs [4], Nishiyama [5], Borgers and Burgers [15] and Jaswon and Wheeler [28]. Further works are required to validate the interest of the approach, but the fact that the KS distortion matrix "predicts" the $(\bar{1}11)_\gamma$ and $(\bar{2}25)_\gamma$ habit planes is very encouraging.

**Table 1** Characteristics of the distortions associated to Bain, Pitsch and Kurdjumov-Sachs (KS) orientation relationships.

| | Final orientation relationship (OR) | | |
|---|---|---|---|
| | **Bain** | **Pitsch** | **KS** |
| **Components of the coordinate transformation matrices** $\mathbf{T}_{\alpha\gamma}$ | | | |
| $[\mathbf{B}_0^\gamma \to \mathbf{B}_c^\gamma] =$ $[\mathbf{x}^\gamma, \mathbf{y}^\gamma, \mathbf{z}^\gamma]$ | $\begin{bmatrix} 1 & 0 & 0 \\ 0 & 1 & 0 \\ 0 & 0 & 1 \end{bmatrix}$ | $\begin{bmatrix} 1/\sqrt{2} & -1/\sqrt{2} & 0 \\ 1/\sqrt{2} & 1/\sqrt{2} & 0 \\ 0 & 0 & 1 \end{bmatrix}$ | $\begin{bmatrix} 1/\sqrt{2} & 1/\sqrt{6} & 1/\sqrt{3} \\ 1/\sqrt{2} & -1/\sqrt{6} & -1/\sqrt{3} \\ 0 & 2/\sqrt{6} & -1/\sqrt{3} \end{bmatrix}$ |
| $[\mathbf{B}_0^\alpha \to \mathbf{B}_c^\alpha] =$ $[\mathbf{x}^\alpha, \mathbf{y}^\alpha, \mathbf{z}^\alpha]$ | $\begin{bmatrix} 1 & 0 & 0 \\ 0 & 1/\sqrt{2} & -1/\sqrt{2} \\ 0 & 1/\sqrt{2} & 1/\sqrt{2} \end{bmatrix}$ | $\begin{bmatrix} 1/\sqrt{3} & 1/\sqrt{6} & -1/\sqrt{2} \\ 1/\sqrt{3} & 1/\sqrt{6} & 1/\sqrt{2} \\ 1/\sqrt{3} & -2/\sqrt{6} & 0 \end{bmatrix}$ | $\begin{bmatrix} 1/\sqrt{3} & 1/\sqrt{6} & -1/\sqrt{2} \\ 1/\sqrt{3} & 1/\sqrt{6} & 1/\sqrt{2} \\ 1/\sqrt{3} & -2/\sqrt{6} & 0 \end{bmatrix}$ |
| **Rotational part of the coordinate transformation matrices for directions and planes** | | | |
| $\mathbf{R}_{\alpha\gamma}$ | $\begin{bmatrix} 1 & 0 & 0 \\ 0 & 1/\sqrt{2} & -1/\sqrt{2} \\ 0 & 1/\sqrt{2} & 1/\sqrt{2} \end{bmatrix}$ | $\begin{bmatrix} \frac{\sqrt{2}-1}{2\sqrt{3}} & \frac{\sqrt{2}+1}{2\sqrt{3}} & \frac{-1}{\sqrt{2}} \\ \frac{\sqrt{2}-1}{2\sqrt{3}} & \frac{\sqrt{2}+1}{2\sqrt{3}} & \frac{1}{\sqrt{2}} \\ \frac{\sqrt{2}+2}{2\sqrt{3}} & \frac{\sqrt{2}-2}{2\sqrt{3}} & 0 \end{bmatrix}$ | $\begin{bmatrix} 1/6 & \sqrt{2/3}-1/6 & 1/\sqrt{6}+1/3 \\ \sqrt{2/3}+1/6 & -1/6 & -1/\sqrt{6}+1/3 \\ 1/\sqrt{6}-1/3 & 1/\sqrt{6}+1/3 & -2/3 \end{bmatrix}$ |
| **Distortion matrices for directions** | | | |
| $\mathbf{D}_0^{OR} =$ | $\begin{bmatrix} \sqrt{2/3} & 0 & 0 \\ 0 & 2/\sqrt{3} & 0 \\ 0 & 0 & 2/\sqrt{3} \end{bmatrix}$ | $\begin{bmatrix} \frac{1+\sqrt{2}}{3} & \frac{2-\sqrt{2}}{3} & 0 \\ \frac{1-\sqrt{2}}{3} & \frac{2+\sqrt{2}}{3} & 0 \\ 0 & 0 & 2/\sqrt{3} \end{bmatrix}$ | $\begin{bmatrix} \sqrt{6}/18+2/3 & -\sqrt{6}/18+1/3 & \sqrt{6}/6-1/3 \\ -\sqrt{6}/18 & \sqrt{6}/18+1 & -\sqrt{6}/6+1/3 \\ \sqrt{6}/9-1/3 & -\sqrt{6}/9+1/3 & \sqrt{6}/3+1/3 \end{bmatrix}$ |
| **Distortion matrices for planes** | | | |
| $^T(\mathbf{D}_0^{OR})^{-1} =$ | $\begin{bmatrix} \sqrt{3/2} & 0 & 0 \\ 0 & \sqrt{3}/2 & 0 \\ 0 & 0 & \sqrt{3}/2 \end{bmatrix}$ | $\begin{bmatrix} \frac{1+\sqrt{2}}{2} & \frac{2-\sqrt{2}}{4} & 0 \\ \frac{1-\sqrt{2}}{2} & \frac{2+\sqrt{2}}{4} & 0 \\ 0 & 0 & \sqrt{3}/2 \end{bmatrix}$ | $\begin{bmatrix} \sqrt{6}/12+1 & -\sqrt{6}/24+1/4 & \sqrt{6}/8-1/4 \\ -\sqrt{6}/12 & \sqrt{6}/24+3/4 & -\sqrt{6}/8+1/4 \\ \sqrt{6}/6-1/2 & -\sqrt{6}/12+1/4 & \sqrt{6}/4+1/4 \end{bmatrix}$ |
| **Images of directions** | | | |
| 3 $<100>_\gamma$ | | 1 $<100>_\alpha$    2 $<110>_\alpha$ | |
| 6 $<110>_\gamma$ | | 4 $<111>_\alpha$    2 $<100>_\alpha$ | |
| 4 $<111>_\gamma$ | | 4 $<210>_\alpha$ | |
| 12 $<112>_\gamma$ | | 8 $<113>_\alpha$    4 $<110>_\alpha$ | |
| **Images of planes** | | | |
| 3 $\{100\}_\gamma$ | | 1 $\{100\}_\alpha$    2 $\{110\}_\alpha$ | |
| 6 $\{110\}_\gamma$ | | 4 $\{211\}_\alpha$    2 $\{100\}_\alpha$ | |
| 4 $\{111\}_\gamma$ | | 4 $\{110\}_\alpha$ | |
| 12 $\{112\}_\gamma$ | | 8 $\{123\}_\alpha$    4 $\{120\}_\alpha$ | |

**Acknowledgements**   I am particularly indebted to thank Prof. Roland Logé for the opportunity he gave me to continue metallurgy at the laboratory of thermomechanical metallurgy (LMTM). I also would like to thank PX Group for their generous support to the laboratory. I am grateful to Prof. Michel Perez (INSA-Lyon) for our discussions and his first molecular dynamic calculations. More works are required before publication, but the first results are very interesting. I thank the reviewers who corrected some nomenclature errors and tried to moderate my point of view on PTMC.

**FIGURES CAPTIONS**

Fig. 1. *Schematic position of the present work. (a) PTMC combines coherently the lattice and shape deformation to predict the orientation relationships and shapes. Our approach only tries to establish a link between the atoms and the lattice deformations.(b) Brief summary of the inputs and outputs of our work and its limitation in comparison with PTMC.*

Fig. 2. *Coordinate transformation and distortion matrices. (a) Correspondence between two bases of same crystal. (b) Correspondence between the bases of crystals of different phases. (c) Transformation by rotation of an atomic row (angular distortive transformation). (d) Shear transformation.*

Fig. 3. *Schematic views on the $(001)_\gamma$ plane of the fcc-bcc distortion associated to (a) Bain and (b) Pitsch OR. The initial $\gamma$ fcc phase is in blue and the final $\alpha$ bcc phase is in red. The compression axis is **a**. This axis is represented along the horizontal direction for Bain and rotated by 45° for Pitsch for practical graphical reasons. In the Bain distortion, the direction **a** is unrotated. In the Pitsch distortion, the direction **x**=**a**+**b** is unrotated and undistorted and noted **N** for neutral.*

Fig. 4. *Schematic view of the fcc crystal, (a) in 3D and (b) in a flatten representation of the face centred cube. The distortion of the **a** and **b** axes in the $(001)_\gamma$ plane implies a distortion along the **c**= $[001]_\gamma$ direction due to the hard sphere packing.*

Fig. 5. *Compression and dilatation along the PM = $[100]_\gamma$ and MQ = $[010]_\gamma$ axes of the $(001)_\gamma$ face during the Pitsch distortion. **PQ** = $[110]_\gamma$= $[111]_\alpha$ = **N***

Fig. 6. *3D scheme of the primitive basis $B_p$ = (**x, y, z**) basis used to calculate the distortion leading to the KS OR. (a) Fcc cube with $(\bar{1}11)_\gamma$ plane marked by the PQR triangle; **x** // P0 = ½ $[110]_\gamma$, **y** // PK = ½ $[101]_\gamma$, **z** = PM = $[100]_\gamma$. (b) 3D representation of the $(\bar{1}11)_\gamma$ plane. The iron atom in M moves such that the distance PO = PK = MO =MK remains constant. The orthonormal basis $B_s$ = ($x_s, y_s, z_s$) associated to $B_p$ is also indicated.*

Fig. 7. *3D representation with VPython of the fcc-bcc distortion associated to Bain, Pitsch and KS ORs. The initial fcc crystal (8x8 cells) is in blue, with the **X, Y** and **Z** directions along the $[100]_\gamma$, $[010]_\gamma$ and $[001]_\gamma$ directions. With Bain, the crystal is compressed along **X** and extended along **Y** and **Z**, and these directions are unrotated. With Pitsch, the direction **Z** is unrotated and the direction **X**+**Y** = $[110]_\gamma$ (noted **N** for neutral) is invariant. With KS, the **X, Y** and **Z** directions are rotated but the direction **N** = **X**+**Y** is invariant and the $(\bar{1}11)_\gamma$ plane is unrotated. The fcc crystals after distortions are in red. The iron atoms inside the distorted fcc crystals form a bcc crystal.*

Fig. 8. *Proposition of experiment to determine the natural fcc-bcc martensitic distortion. A small cube (few μm³) of metastable austenite is prepared by FIB and deposited at the tip of sharp needle. An absolute reference frame is engraved at the surface of the tip. Then the tip is cooled below room temperature such that the austenite is transformed into only one variant of martensite (due to its small size). If the cube is not rotated but just squeezed along a <100> direction, then the natural distortion is Bain. If the cube is rotated the natural distortion is not Bain. It can be Pitsch or KS depending on lines and planes that are let invariant.*

Fig. 9. *Distortion of the $(\bar{1}11)_\gamma$ plane into a $(\bar{1}10)_\alpha$ plane. At the top, 3D view by orienting the fcc crystal along the $[\bar{1}11]_\gamma$ axis. The planes are delineated by the yellow lines. At the bottom, 2D representation of the surface change $S_{bcc}/S_{fcc} = \sin(70.5°)/\sin(60°) \approx 1.088$. The distance d is the diameter of the iron atoms.*

Fig. 10. *Modified one-step model. In Fig. 8 of ref. [18], the distortion was supposed to lead to Pitsch OR, whereas in the present version, the distortion leads to the KS OR. Rotations A and B, and the associated Pitsch and NW ORs, result from the strain field generated by the distortion.*

Fig. 11. *Homogeneous transformation of an atomic plane. The initial plane (at the left side) given by the reciprocal vector **g** (in blue) is transformed into distorted plane **g'**. The planar change Δ**g** = **g'**-**g** can be decomposed into two parts: Δ**g**$_{//}$ and Δ**g**$_\perp$. They represent the change of the interplanar distance (at the top right corner) and the planar tilt (at the bottom right corner), respectively.*

Fig. 12. *Graphical representations with Mathematica of ∥ Δ**g**$_\perp$ ∥ given in %. (a) Spherical 3D representation of ∥ Δ**g**$_\perp$ ∥ for normalized reciprocal vectors **g**, (b) surface representation of ∥ Δ**g**$_\perp$ ∥ according to the spherical coordinates θ and φ, with θ ∈ [0,π] and θ ∈ [0,2π]. (c) 2D representation with contours. (d) Enlargement of the region around the two local minima $(\bar{1}11)_\gamma$ and $(\bar{2}25)_\gamma$.*

# FIGURES

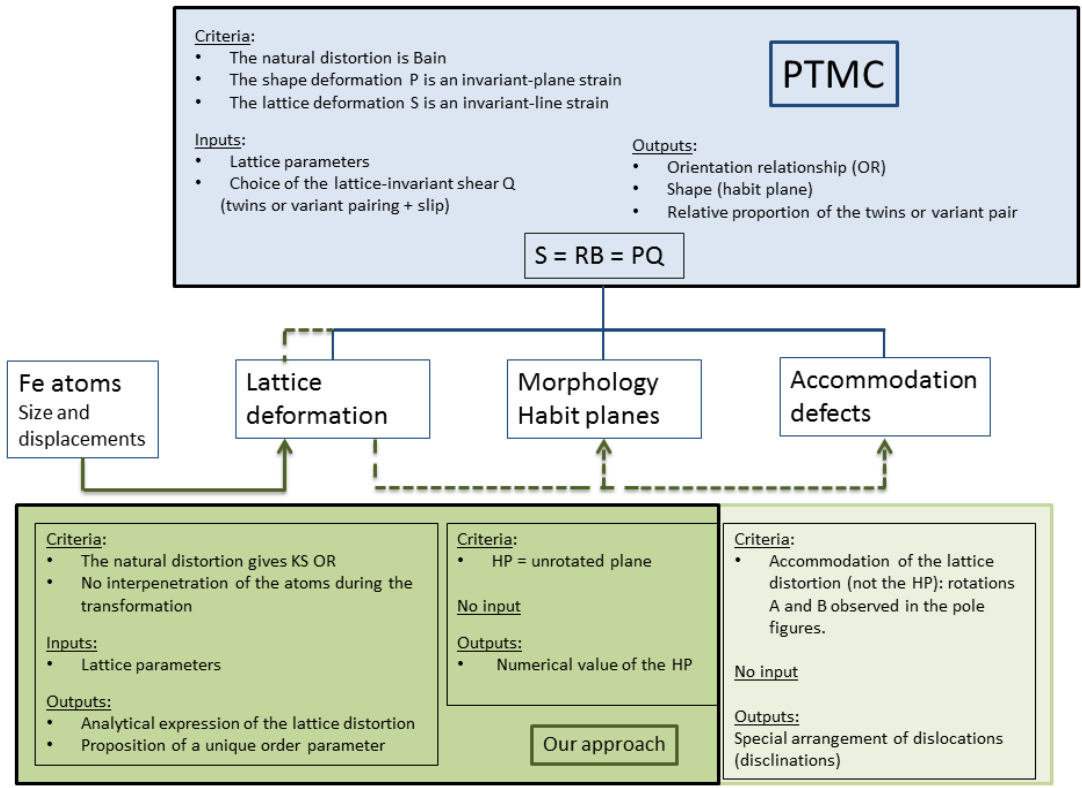

Fig. 1. *Schematic position of the present work. (a) PTMC combines coherently the lattice and shape deformation to predict the orientation relationships and shapes. Our approach only tries to establish a link between the atoms and the lattice deformations.(b) Brief summary of the inputs and outputs of our work and its limitation in comparison with PTMC.*

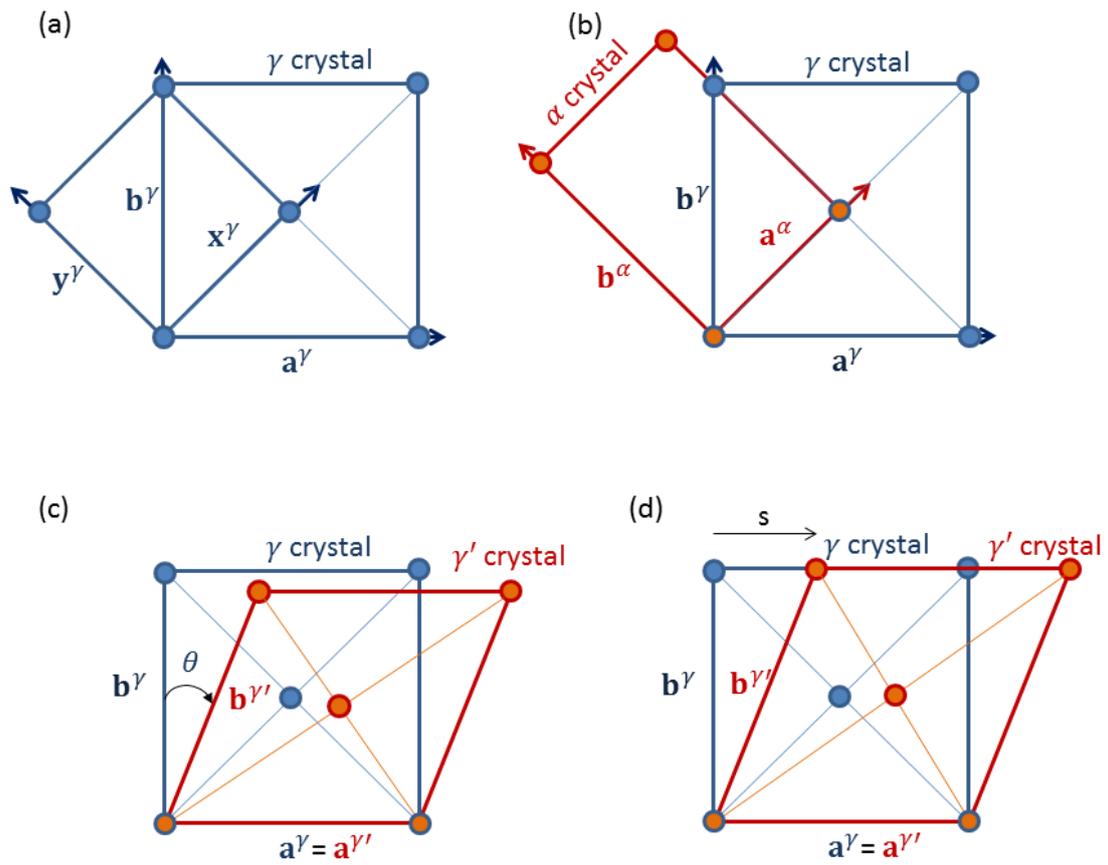

Fig. 2. *Coordinate transformation and distortion matrices. (a) Correspondence between two bases of same crystal. (b) Correspondence between the bases of crystals of different phases. (c) Transformation by rotation of an atomic row (angular distortive transformation). (d) Shear transformation.*

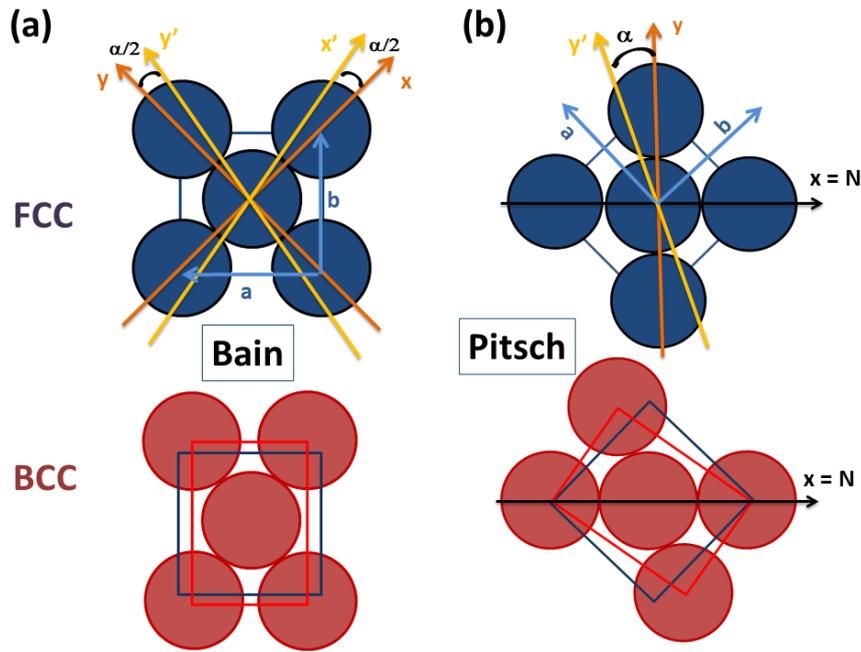

Fig. 3. *Schematic views on the (001)$_\gamma$ plane of the fcc-bcc distortion associated to (a) Bain and (b) Pitsch OR. The initial $\gamma$ fcc phase is in blue and the final $\alpha$ bcc phase is in red. The compression axis is **a**. This axis is represented along the horizontal direction for Bain and rotated by 45° for Pitsch for practical graphical reasons. In the Bain distortion, the direction **a** is unrotated. In the Pitsch distortion, the direction **x**=**a**+**b** is unrotated and undistorted and noted **N** for neutral.*

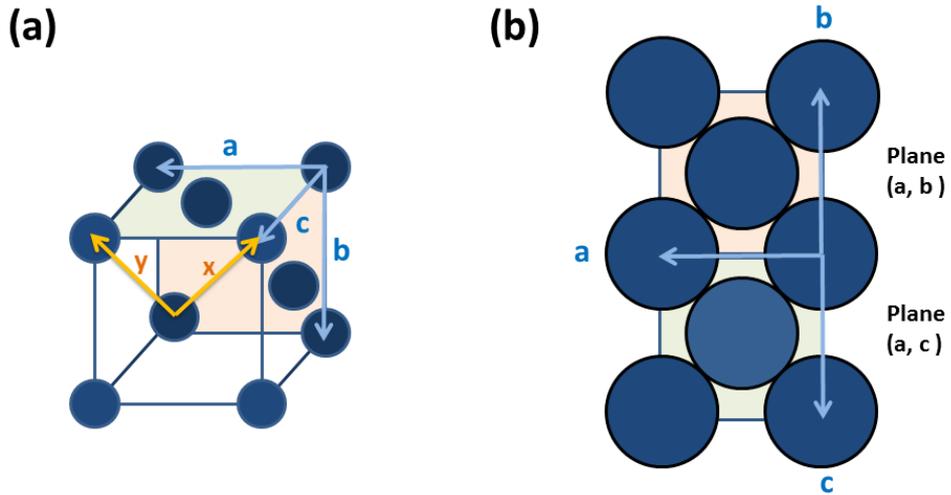

Fig. 4. *Schematic view of the fcc crystal, (a) in 3D and (b) in a flatten representation of the face centred cube. The distortion of the **a** and **b** axes in the (001)$_\gamma$ plane implies a distortion along the **c** = [001]$_\gamma$ direction due to the hard sphere packing.*

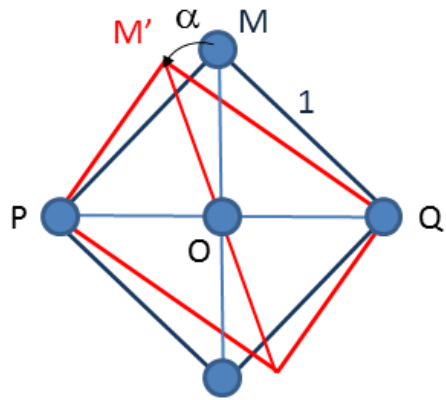

Fig. 5. *Compression and dilatation along the **PM** = [100]$_\gamma$ and **MQ** = [010]$_\gamma$ axes of the (001)$_\gamma$ face during the Pitsch distortion. **PQ** = [110]$_\gamma$ = [111]$_\alpha$*

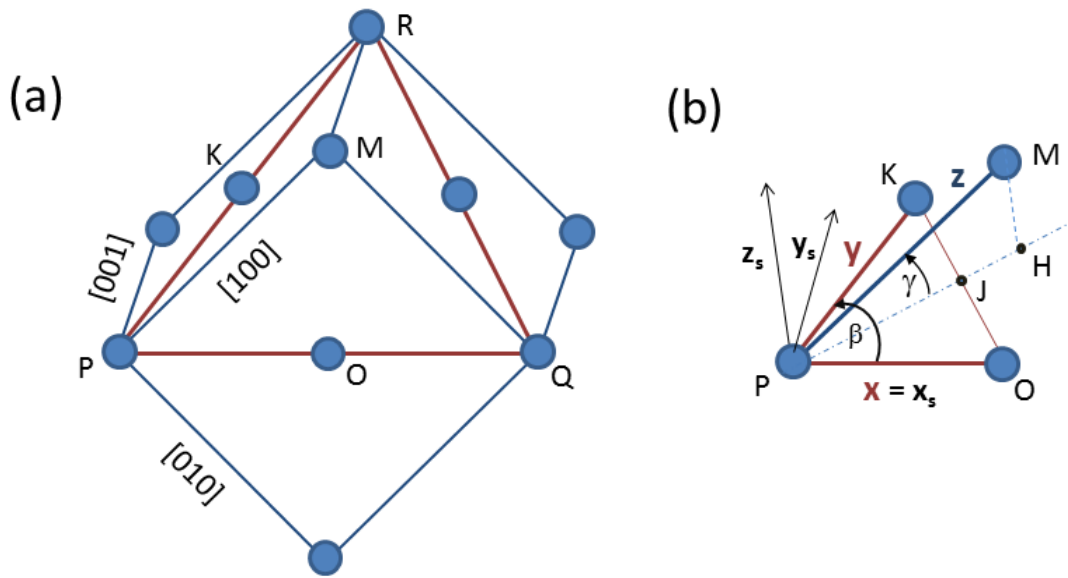

Fig. 6. *3D scheme of the primitive basis $B_p = (x, y, z)$ basis used to calculate the distortion leading to the KS OR. (a) Fcc cube with $(\bar{1}11)_\gamma$ plane marked by the PQR triangle; $x \mathbin{/\mkern-2mu/} P0 = \frac{1}{2}[110]_\gamma$, $y \mathbin{/\mkern-2mu/} PK = \frac{1}{2}[101]_\gamma$, $z = PM = [100]_\gamma$. (b) 3D representation of the $(\bar{1}11)_\gamma$ plane. The iron atom in M moves such that the distance PO = PK = MO = MK remains constant. The orthonormal basis $B_s = (x_s, y_s, z_s)$ associated to $B_p$ is also indicated.*

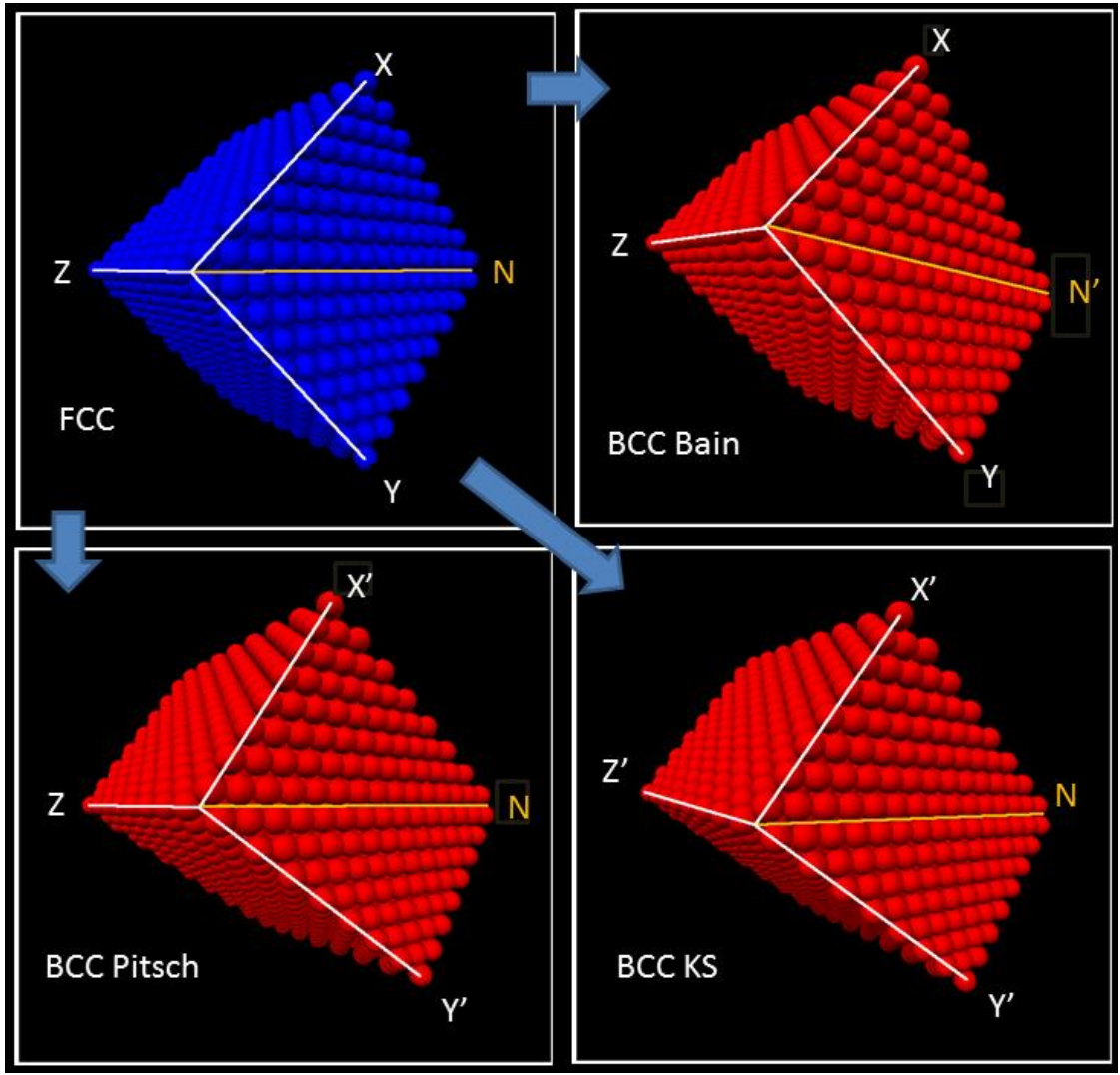

Fig. 7. *3D representation with VPython of the fcc-bcc distortion associated to Bain, Pitsch and KS ORs. The initial fcc crystal (8x8 cells) is in blue, with the **X**, **Y** and **Z** directions along the $[100]_\gamma$, $[010]_\gamma$ and $[001]_\gamma$ directions. With Bain, the crystal is compressed along **X** and extended along **Y** and **Z**, and these directions are unrotated. With Pitsch, the direction **Z** is unrotated and the direction **X**+**Y** = $[110]_\gamma$ (noted **N** for neutral) is invariant. With KS, the **X**, **Y** and **Z** directions are rotated but the direction **N** = **X**+**Y** is invariant and the $(\bar{1}11)_\gamma$ plane is unrotated. The fcc crystals after distortions are in red. The iron atoms inside the distorted fcc crystals form a bcc crystal.*

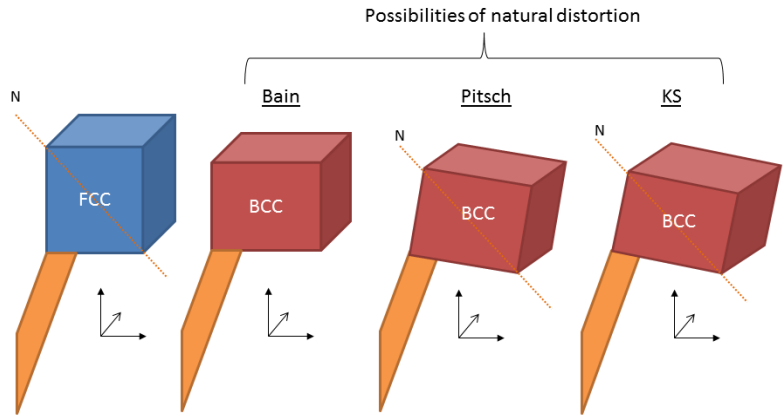

Fig. 8. *Proposition of experiment to determine the natural fcc-bcc martensitic distortion. A small cube (few µm³) of metastable austenite is prepared by FIB and deposited at the tip of sharp needle. An absolute reference frame is engraved at the surface of the needle. Then the cube is cooled below room temperature so that the austenite is transformed into only one variant of martensite (due to its small size). If the cube is not rotated but just squeezed along a <100> direction, then the natural distortion is Bain. If the cube is rotated the natural distortion is not Bain. It can be Pitsch or KS depending on lines and planes that are let invariant.*

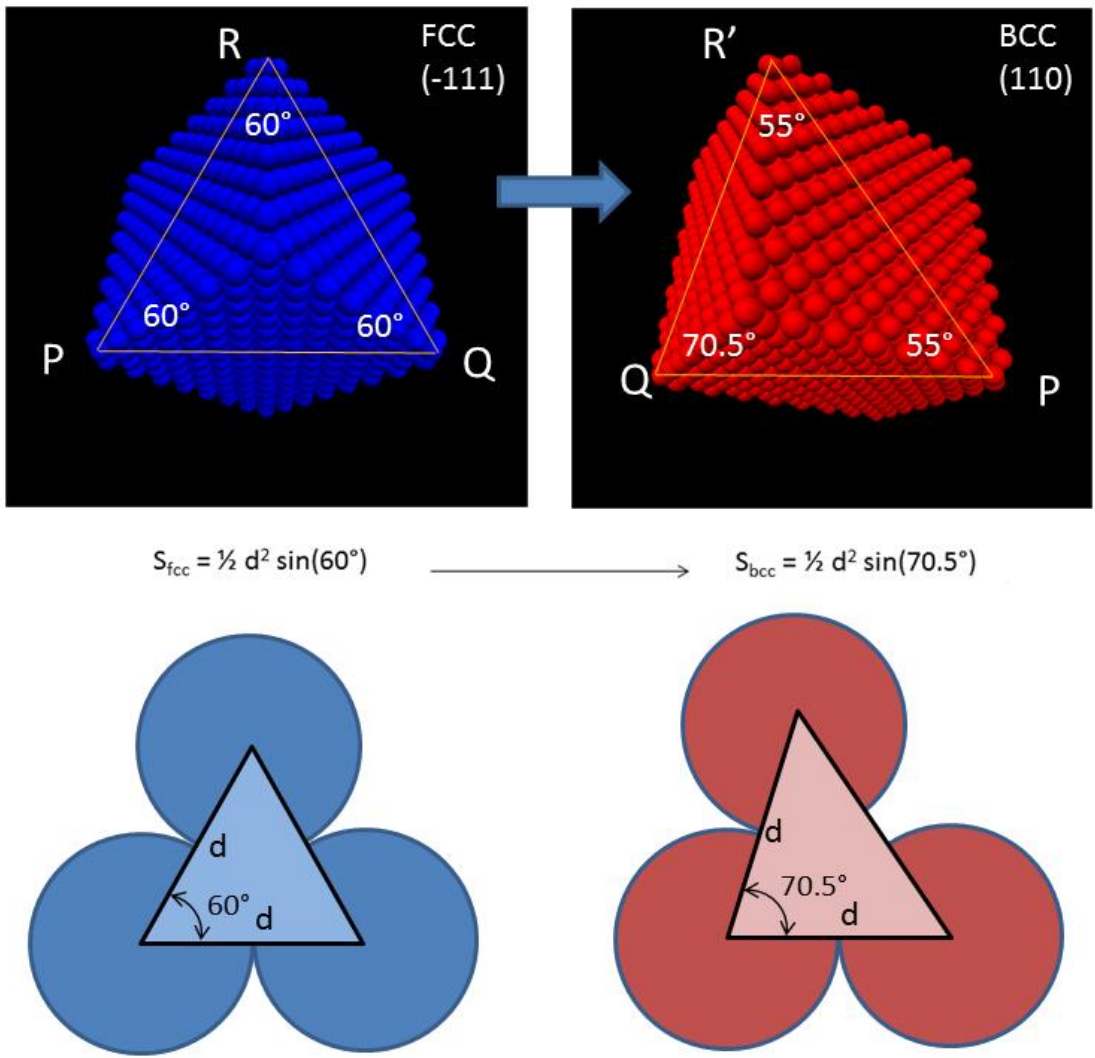

Fig. 9. *Distortion of the $(\bar{1}11)_\gamma$ plane into a $(\bar{1}10)_\alpha$ plane. At the top, 3D view by orienting the fcc crystal along the $[\bar{1}11]_\gamma$ axis. The planes are delineated by the yellow lines. At the bottom, 2D representation of the surface change $S_{bcc}/S_{fcc} = sin(70.5°)/sin(60°) \approx 1.088$. The distance d is the diameter of the iron atoms.*

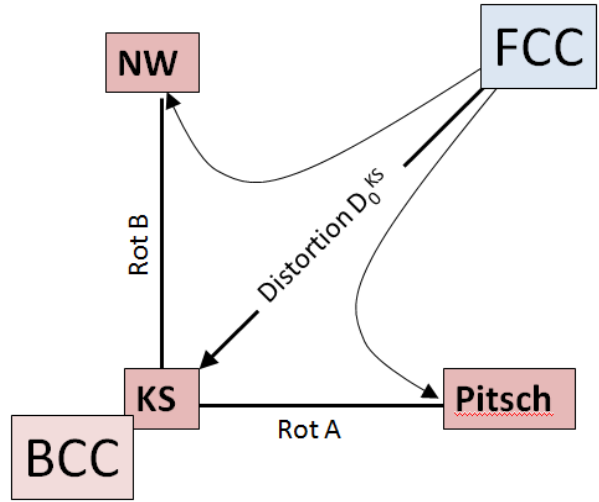

Fig. 10. *Modified one-step model. In Fig. 8 of ref. [18], the distortion was supposed to lead to Pitsch OR, whereas in the present version, the distortion leads to the KS OR. Rotations A and B, and the associated Pitsch and NW ORs, result from the strain field generated by the distortion.*

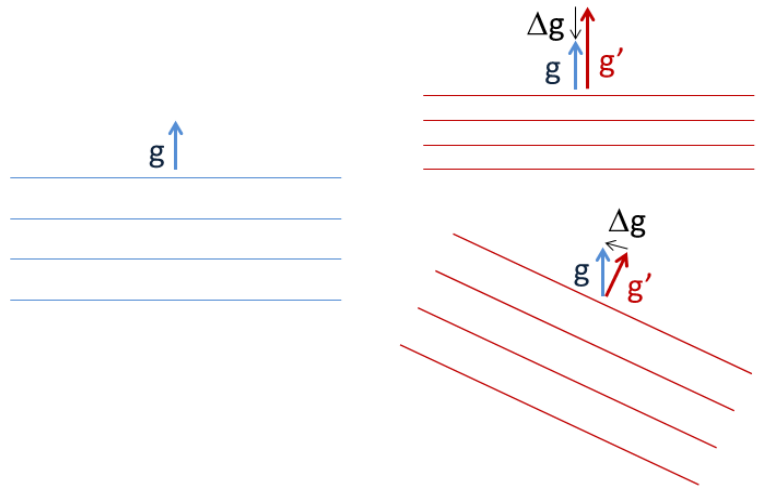

Fig. 11. *Homogeneous transformation of an atomic plane. The initial plane (at the left side) given by the reciprocal vector **g** (in blue) is transformed into a distorted plane **g'**. The planar change $\Delta g = g'-g$ can be decomposed into two parts: $\Delta g_{//}$ and $\Delta g_{\perp}$. They represent the change of the interplanar distance (at the top right corner) and the planar tilt (at the bottom right corner), respectively.*

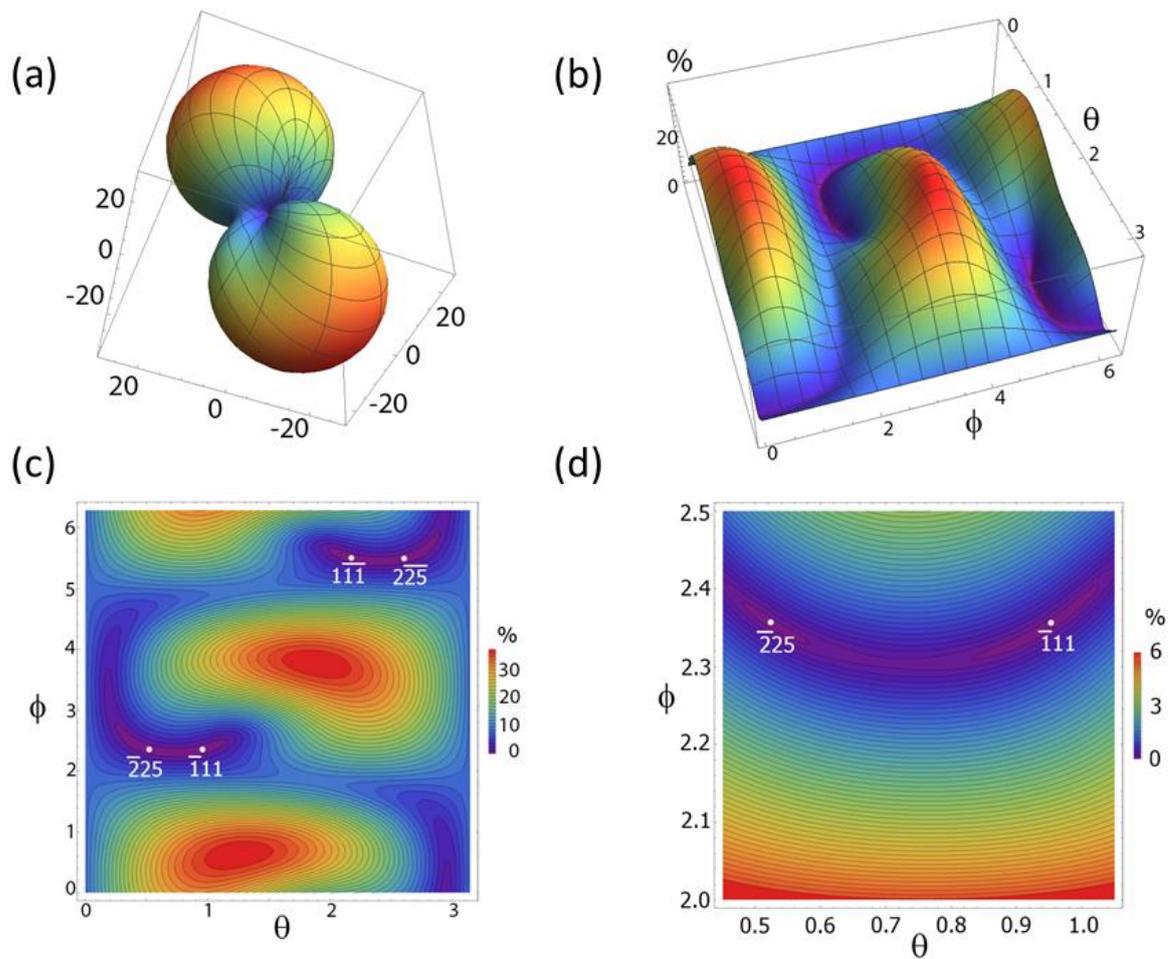

Fig. 12. *Graphical representations with Mathematica of $\|\Delta g_\perp\|$ given in %. (a) Spherical 3D representation of $\|\Delta g_\perp\|$ for normalized reciprocal vectors **g**, (b) surface representation of $\|\Delta g_\perp\|$ according to the spherical coordinates θ and φ, with θ ∈ [0,π] and θ ∈ [0,2π]. (c) 2D representation with contours. (d) Enlargement of the region around the two local minima $(\bar{1}11)_\gamma$ and $(\bar{2}25)_\gamma$.*